\documentclass{article}

\def\giorno{3/10/2023}

\usepackage{color}
\usepackage{amsmath,amsfonts,amssymb}
\usepackage{cite}

\def\gcite#1{{\cite{#1}}}

\makeatletter
\@addtoreset{equation}{section}
\makeatother

\def\a{\alpha}
\def\b{\beta}
\def\ga{\gamma}
\def\de{\delta}   
\def\eps{\varepsilon}
\def\vphi{\varphi}
\def\la{\lambda}

\def\s{\sigma}

\def\vphi{\varphi}

\def\De{\Delta}

\def\pa{\partial}

\def\xb{{\bf x}}

\def\wb{{\bf w}}

\def\o+{\oplus}

\def\<{\langle}
\def\>{\rangle}

\def\({\left(}
\def\){\right)}
\def\[{\left[}
\def\]{\right]}
\def\=#1{\bar #1}
\def\~#1{\widetilde #1}
\def\wt#1{\widetilde #1}
\def\.#1{\dot #1}
\def\^#1{\widehat #1}

\def\"#1{\ddot #1}

\def\eeq{\end{equation}}
\def\beq{\begin{equation}}

\def\beql#1{\begin{equation} \label{#1}}

\def\eqref#1{(\ref{#1})}

\def\EOR{ \hfill $\odot$ \medskip}
\def\EOE{ \hfill $\diamondsuit$ \medskip}

\def\symmref{AVL,CGbook,KrV,Olver1,Olver2,Stephani}
\def\sderef{Arnold,Evans,Fre,Ikeda,Kampen,Oksendal,Stroock,Amir}

\begin{document}

\title{On the integration of Ito equations with a random or a W-symmetry}

\author{G. Gaeta$^{1,2}$\thanks{Also member of GNFM-INdAM and of INFN (sezione di Milano, MMNLP project)} {$\ $}\thanks{giuseppe.gaeta@unimi.it} \\
${}^1$ {\it Dipartimento di Matematica, Universit\`a degli Studi
di
Milano,} \\ {\it via Saldini 50, 20133 Milano (Italy)} \\
${}^2$ {\it SMRI, 00058 Santa Marinella (Italy)} }
\date{\giorno}


\maketitle

\begin{abstract}
\noindent Symmetries can be used to integrate scalar Ito equation -- or reduce systems of such equations -- by the Kozlov substitution, i.e. passing to symmetry adapted coordinates. While the theory is well established for so called deterministic standard symmetries (the class originally studied by Kozlov), some points need clarification for so called random standard symmetries and W-symmetries. This paper is devoted to such clarification; in particular we note that the theory naturally calls, for these classes of symmetries, to also consider generalized Ito equations; and that while Kozlov theory is extended substantially unharmed for random standard symmetries, W-symmetries should be handled with great care, and cannot be used towards integration of stochastic equations, albeit they have different uses.
\end{abstract}



\newpage

\section{Introduction}

Symmetry reduction is a standard tool in the study of nonlinear deterministic differential equations \gcite{\symmref}; it entered in the toolbox of those studying stochastic differential equations (SDEs in the following) \gcite{\sderef} thanks to the work of R. Kozlov in 2010 \gcite{Koz1,Koz2,Koz3,Koz2018}. In fact, he determined the change of random variable allowing to take full advantage of the existence of a symmetry in the case of a SDE, pretty much as in the case of deterministic ODEs and PDEs, where symmetry is used through the introduction of \emph{symmetry adapted variables} \gcite{\symmref}. In the case of a scalar SDE, determination of a symmetry (of appropriate form, see below) leads to its explicit integration.

Kozlov's result provided a powerful motivation for the study of symmetries of SDEs; in following years the class of symmetries considered in the literature was substantially extended. In particular, two major classes of symmetries are by now usually considered besides those originally considered by Kozlov, dubbed in contemporary nomenclature \emph{deterministic standard symmetries}. These other classes are \emph{random standard symmetries} \gcite{GS17,Koz18a,Koz18b} on the one hand, and \emph{W-symmetries} \gcite{GSW} on the other (see below for a precise definition).

A critical analysis of the literature shows that while the situation is very well clarified for deterministic standard symmetries, and a number of results are available for random ones \gcite{Koz18a}, still some relevant results are not clearly established for random standard symmetries, and the role of W-symmetries in SDE reduction or integration is on an even more shaky basis.

The purpose of this paper is to clarify the situation in this regard. We anticipate that on the one hand in the case of random deterministic symmetries we will show that they are ``as good as deterministic standard symmetries'' for what concerns symmetry reduction and integration of SDEs  (but proving this when multiple reductions are involved requires a substantial extension of the discussion holding for deterministic standard symmetries, as discussed in Sect. \ref{sec:generalized}); on the other hand the use of W-symmetries along the same lines of standard symmetries is alas impossible, contrary to what was maybe hoped when they were introduced \gcite{GSW}. (This does not mean that they are not useful in general; we will discuss this point in Sect.\ref{sec:relW}.)

Roughly speaking, the success of the symmetry approach to SDEs rooted in the use of deterministic standard symmetries is based on two properties of these:
\begin{itemize}
\item Symmetries are preserved under a change of variables;
\item A change of variables which straightens the symmetry vector field preserves the class of Ito equations.
\end{itemize}

As for the first property -- which is obvious for deterministic equations but requires a discussion for stochastic ones -- this holds for random standard symmetries, while it requires additional conditions (which in general are not satisfied) when it comes to considering W-symmetries. We will also see that the second property extends to random standard symmetries provided one accepts to consider \emph{generalized} (rather than proper) Ito equations, while in general it does not hold for W-symmetries. This fact precludes the possibility of using W-symmetries within the Kozlov scheme, at least remaining within the theory of (possibly, generalized) Ito equations.

Our discussion in the main text will only consider scalar equations; this will suffice to make our point, and will avoid unneeded complications (we will slightly deviate from this line of discussion in one Example in Section \ref{sec:relW}). However, the case of higher dimensional systems of Ito equations will be mentioned at several places as a motivation for our study; we will provide the relevant formulas and several Examples of higher dimensional systems in Appendix \ref{app:deteqhigher}.

Several of the Examples presented here (in particular about W-symmetries) are taken from \gcite{GSW}, but considered under a different light.

Summation over repeated indices will always be understood; the end of a Remark or an Example will be marked, respectively, by the symbols $\odot$ and $\diamondsuit$.

\medskip\noindent
{\bf Remark 0.} We should stress two points in our approach which could be confusing and hence needs to be clarified. 

First, as mentioned above, our approach will consider changes of variables in Ito equations; these will of course be performed according to the rules of stochastic (Ito) calculus. On the other hand, we will consider symmetries described by the action of a vector field in the $({\bf x},t; {\bf w})$ space (here $w^i$ are the Wiener processes driving the Ito equation).  Vector fields describe diffeomorphisms, and under a change of variables their  components change under the familiar chain rule. Thus we will have two different rules governing the behavior of equations and vector fields under the considered change of variables -- which, as we will seen in a moment, will \emph{not} be the most general smooth changes of variables in $({\bf x},t; {\bf w})$ space. In other words, when considering vector fields we will look at the $w$ variables on the same footing as the $x$ and $t$ ones; it will be only when discussing Ito equations that we take into account that the time evolution of $w = w(t)$ is non-differentiable, and that Ito calculus should come into play.

Second, we attempt at reproducing the approach -- and as far as possible the results -- obtained by symmetry theory in the case of deterministic differential equations (which yields a very transparent construction). A different approach to symmetry of stochastic equations, also allowing for more general changes of variables, could be followed and has indeed been considered in the literature. In this case one does not need to consider vector fields and should instead remain within the standard frame of stochastic calculus; in particular, the effect of (fully general) changes of variables should be taken into account by the classical Girsanov formula. We refer to \cite{DVMU1,DVMU2} for this approach; see also \cite{PrZ,ChT,DVMU3,DVMU4,DVMU5,DVMU6}. An overall discussion of symmetries of stochastic equations also putting this approach in perspective is provided in \cite{ADV}. \EOR

\section{Proper and generalized Ito equations; and why the latter are relevant in multiple symmetry reduction of the former}
\label{sec:generalized}

In the following, we will be naturally led to consider, besides \emph{proper} Ito equations
\beql{eq:Ito} dx \ = \ f(x,t) \ dt \ + \ \s (x,t) \ d w \ , \eeq also \emph{generalized} Ito equations, i.e. stochastic differential equations of the form
\beql{eq:gIto} dx \ = \ f (x,t;w) \ dt \ + \ \s (x,t;w) \ dw \ ; \eeq
here and above, $w = w(t)$ is a Wiener process, also called the \emph{driving process} for the equation at hand.

We will consider \emph{symmetries} of SDEs; by this we always mean Lie-point symmetries, generated by a vector field acting in the $(x,t,w)$ space. These will not act on the $t$ variable (thus the distinguished role of time is preserved\footnote{One could consider time reparametrizations \gcite{GRQ1,GRQ2,GGPR}; but this will not change substantially the situation, and produce some complication in formulas. We note in passing that a rescaling of time might be relevant when considering W-symmetries, which produce a rescaling of the noise term.} \gcite{GS17}) and act on the Wiener processes only by reparametrization. In the one-dimensional case these requirements mean the vector field will be of the form
\beql{eq:X} X \ = \ \vphi (x,t,w) \, \pa_x \ + \ r \, w \, \pa_w \ ; \eeq
here $r$ is a real parameter, and $\vphi$ a smooth function. For $r=0$ we speak of \emph{standard symmetries}, while for $r \not= 0$ of (proper) \emph{W-symmetries}. When considering standard symmetries we further discriminate between \emph{deterministic} ones, for which $\vphi = \vphi (x,t)$, and \emph{random} ones, for which $\vphi$ depends effectively on $w$ (in early literature, including \gcite{GSW}, random standard symmetries are called simply ``random symmetries''). In the case of W-symmetries, when $\vphi$ does not depend on $w$ we speak of \emph{split W-symmetries}.

From our point of view, studying symmetries is first of all motivated by the fact that knowing a symmetry -- or more precisely a standard symmetry, see below -- allows to integrate the Ito equation. This is obtained through the \emph{Kozlov substitution}, i.e. passing to the variable
\beq y \ = \ \Phi (x,t,w) \ = \ \int \frac{1}{\vphi (x,t,w) } \ d x  \eeq
while leaving $(t,w)$ unchanged.

A swift computation shows that in the $(y,t,w)$ variables the vector field $X = \vphi \pa_x$ reads
$$ X \ = \ \pa_y \ ; $$ correspondingly the Ito equation is transformed into an equation of the form
\beql{eq_yeq} d y \ = \ F \, dt \ + \ S \, dw \eeq
where $F$ and $S$ can be determined explicitly by Ito calculus (see Remark 3 below for their explicit form).

If we are dealing with a \emph{deterministic} standard symmetry, we get $F = F(t)$, $S = S(t)$, and the transformed equation is hence a new Ito equation, which moreover is immediately and elementarily integrated.

If we are dealing with a \emph{random} standard symmetry, we get $F = F(t,w)$, $S = S(t,w)$, and the transformed equation is hence \emph{not} an Ito equation, but can nevertheless be integrated.

In the case of W-symmetries, this question has -- as far as I know -- not been considered in the literature. In fact, this is one of the questions we will tackle in the present work.

Another problem arises when we consider \emph{systems} of Ito equations; in this case one could have, and try to use towards partial integration, i.e. \emph{reduction}, \emph{multiple} symmetries. We could thus employ one of these symmetries to \emph{reduce} the system (pretty much as we do in the case of symmetric deterministic ODEs \gcite{\symmref}). In order to use a second symmetry to further reduce -- or integrate in the case of a two-dimensional system -- the first reduced system, we should be sure the symmetry is preserved in the reduction process. This in turn raises two kinds of questions:
\begin{itemize}
\item Is a symmetry preserved under a change of variables?
\item Is a symmetry preserved under a symmetry reduction?
\end{itemize}

The second question also arises in dealing with deterministic equations \gcite{\symmref}, and we expect the answer will depend on the Lie algebraic structure of the symmetry algebra; but this matter will not be studied here, and is postponed to future work.

Here we should focus on the first question, which just makes no sense (the answer being trivially positive) in the deterministic context. In fact, when we consider a change of variables, one object, i.e. the vector field, is transformed under the standard chain rule, while the other, i.e. the Ito equation, changes under a different set of rules, i.e. obeying the rules of Ito calculus.\footnote{Actually, one could argue that it makes no sense speaking of symmetries if these are not preserved under a change of variables. So this question is essential to the very existence of symmetries for Ito stochastic equations.}

A general positive answer to this question was given in \gcite{GL1} (see also \gcite{GL2}); but that paper only considered deterministic standard symmetries (this for the simple reason that the other types of symmetries mentioned above had not yet been introduced in the literature). We will thus have to consider and answer this question both for random standard symmetries and for W-symmetries. Obviously, a negative answer would rule out the possibility of using these types of symmetry in multiple reduction.

But, as we have seen above, if the first reduction is made using a \emph{random} symmetry, the transformed equation is not any more a proper Ito equation: it will be instead a generalized one. This calls immediately for consideration of symmetries of generalized Ito equations and for studying their fate under a change of variables (and under a reduction, albeit this makes sense only in higher dimension).  These matters will also be studied in the present paper.

\section{Symmetry of proper and generalized Ito equations. Determining equations}
\label{sec:ito}

Let us first consider the action of a (formal) vector field
\beql{eq:Xdefnew} X \ = \ \vphi (x,t;w) \, \pa_x \ + \ \xi (x,t;w) \, \pa_w \ , \eeq
defined in the $(x,t;w)$ space, on equations of the form \eqref{eq:gIto}, i.e. on a generalized Ito equation (we will see proper Ito equations \eqref{eq:Ito} as a special subclass of these).

The infinitesimal action of $X$ is given by
$$ x \ \to \ x \ + \ \eps  \, \vphi (x,t;w) \ , \ \ \ \ w \ \to \ w \ + \ \eps \, \xi (x,t;w) \ . $$ Applying $X$ on \eqref{eq:gIto} we obtain immediately (see Appendix \ref{app:deteqs} for details) that the equation is invariant if and only if the following \emph{determining equations} are satisfied:
\begin{eqnarray}
\vphi_t \ + \ f \, \vphi_x \ - \ \vphi \, f_x \ + \ \frac12 \, \De (\vphi) &=& \xi \ f_w { \ + \ A (x,t;w)  } \ , \label{eq:fdenew1} \\
\vphi_w \ + \ \s \, \vphi_x \ - \ \vphi \, \s_x \ - \ \s \, \xi_w \  &=& \xi \ \s_w { \ + \ \s^2 \, \xi_x  } \ . \label{eq:fdenew2} \end{eqnarray}
Here and below, $\Delta$ denotes the \emph{Ito Laplacian} \gcite{\sderef}. In the one-dimensional case we are considering throughout the paper, this reads simply
\beql{eq:Delta} \Delta \ = \ \frac{\pa^2}{\pa w^2} \ + \ 2 \ \s \ \frac{\pa^2}{\pa w \pa x} \ + \ \s^2 \ \frac{\pa^2}{\pa x^2} \ . \eeq

Moreover, in \eqref{eq:fdenew1} we have defined, for ease of notation, 
$$ A (x,t;w) \ = \ \s \, f \, \xi_x \ + \ \s \, \xi_t \ + \ \frac12 \, \s \De \xi \ . $$

We are actually interested in the case where
\beq \xi (x,t;w) \ = \ r \ w \eeq
with $r$ a real (possibly zero) constant; see \gcite{GSW} for a discussion in this sense. With this form for $\xi$, we have $A (x,t;w) = 0$. The determining equations for W-symmetries of a generalized Ito equation are then
\begin{eqnarray}
\vphi_t \ + \ f \, \vphi_x \ - \ \vphi \, f_x \ + \ \frac12 \, \De (\vphi) &=& r \, w \ f_w \ , \label{eq:gendeteq1} \\
\vphi_w \ + \ \s \, \vphi_x \ - \ \vphi \, \s_x \ - \ r \, \s  &=& r \, w  \ \s_w \ . \label{eq:gendeteq2}  \end{eqnarray}
Note that for a proper Ito equation, the right hand sides of these just vanish. We recover then the familiar determining equations for W-symmetries of an Ito equation \gcite{GGPR}.

The determining equations for \emph{standard} symmetries are obtained setting $r=0$ in the above.
Note that in this case the equations are exactly the same for proper and generalized Ito equations; they read
\begin{eqnarray}
\vphi_t \ + \ f \, \vphi_x \ - \ \vphi \, f_x \ + \ \frac12 \, \De (\vphi) &=& 0 \ , \label{eq:deteqstand1} \\
\vphi_w \ + \ \s \, \vphi_x \ - \ \vphi \, \s_x  &=& 0 \ . \label{eq:deteqstand2} \end{eqnarray}

\section{The Stratonovich counterpart}
\label{sec:strat}

As well known \gcite{\sderef}, to a (proper) Ito equation \eqref{eq:Ito} is associated a \emph{Stratonovich} equation\footnote{The Ito-Stratonovich correspondence does actually present some subtle points; for these the reader is referred to \gcite{Stroock}.}
\beql{eq:Strat} d x \ = \ b (x,t) \, dt \ + \ \s (x,t) \circ dw \ ; \eeq the drift coefficient $f(x,t)$ of the Ito equation \eqref{eq:Ito} and that of the associated Stratonovich one \eqref{eq:Strat}, i.e. the \emph{Stratonovich drift} $b(x,t)$ -- are related by the \emph{Stratonovich map} \beql{eq:Smap} b (x,t) \ = \ f(x,t) \ - \ \frac12 \, \s (x,t) \, \s_x (x,t) \ . \eeq

In the following it will be convenient to also consider the Stratonovich counterparts of our equations.

We may note that as here we are also considering generalized Ito equations \eqref{eq:gIto}, we should as well consider \emph{generalized Stratonovich equations}, i.e. equations of the form
\beql{eq:gStrat} d x \ = \ b (x,t,w) \, dt \ + \ \s (x,t,w) \circ dw \ . \eeq
In particular, we associate to the generalized Ito equation \eqref{eq:gIto} the equation \eqref{eq:gStrat} with the same noise coefficient $\s (x,t,w)$ and with the drift coefficients $f$ and $b$ being related by the \emph{generalized Stratonovich map} \beql{eq:gSmap} b (x,t,w) \ = \ f(x,t,w) \ - \ \frac12 \, \[ \s (x,t,w) \, \s_x (x,t,w) \ + \ \s_w (x,t,w) \] \ . \eeq This formula is derived in Appendix \ref{app:relationseqs}.

Proceeding as for Ito equations (see again Appendix \ref{app:deteqs} for details), the determining equations for symmetries of the generalized Stratonovich equation \eqref{eq:gStrat} are
\begin{eqnarray}
\vphi_t \ + \ b \, \vphi_x \ - \ \vphi \, b_x  &=& r \, w \ b_w \ ,  \label{eq:deteqstrat1} \\
\vphi_w \ + \ \s \, \vphi_x \ - \ \vphi \, \s_x \ - \ r \, \s  &=& r \, w  \ \s_w \ . \label{eq:deteqstrat2} \end{eqnarray}
Note that the second equation \eqref{eq:deteqstrat2} is just the same as the corresponding one for the Ito equation, \eqref{eq:gendeteq2}. This is not the case if we compare the first determining equations \eqref{eq:deteqstrat1} and \eqref{eq:gendeteq1} in themselves; however these could and should be thought as defined on the space of solutions to the second determining equation. A detailed discussion in this regard will be given in the following, see Section \ref{sec:relation}.

The determining equations for \emph{standard} symmetries are again obtained setting $r=0$ in the above; in particular, the r.h.s. of both equations are then zero.
Note also that in this case (as also occurs for Ito equations) the equations are exactly the same for proper and generalized Stratonovich equations; they read
\begin{eqnarray}
\vphi_t \ + \ b \, \vphi_x \ - \ \vphi \, b_x &=& 0 \ , \label{eq:deteqstratstand1} \\
\vphi_w \ + \ \s \, \vphi_x \ - \ \vphi \, \s_x  &=& 0 \ . \label{eq:deteqstratstand2} \end{eqnarray}

\section{The vector fields point of view. Straightening}
\label{sec:straight}

When we study symmetry of SDE, we have to consider two kind of objects, with rather different properties: \emph{vector fields}, and the \emph{Ito equation} itself. They behave differently under a change of variable: vector fields transform according to the usual properties (that is, the chain rule), while Ito equations not; in particular, differentials transform according to Ito calculus.\footnote{This could be circumvented passing to consider the \emph{Stratonovich} equation associated to the given Ito one. One should again remember that the correspondence between Ito and Stratonovich equations involve some subtleties \gcite{Stroock}, and this is one of the reasons why we prefer to deal with the Ito equation itself. We also recall that correspondence between (determining equations, and hence) symmetries of an Ito equation and the associated Stratonovich one has been proved in full generality for \emph{standard} symmetries in \gcite{GL1}. That proof does not apply to W-symmetries (which had not been introduced in the literature at the time); see in this respect the discussion below in Section \ref{sec:persW}.}

Pretty much in the same way as when dealing with deterministic equations, once we have determined a symmetry it is convenient to pass to \emph{symmetry-adapted variables}. These are obtained by \emph{straightening} the vector field \gcite{ArnODE,ArnodeS}.

In fact, the Kozlov substitution is exactly the change of variables realizing the map to symmetry adapted variables.

\subsection{Standard symmetries}

The kind of vector fields which are mainly useful in considering symmetry (and integration) of SDEs corresponds to \emph{standard symmetries} and is of the form \beql{eq:Xst} X \ = \ \vphi (x,t;w) \, \pa_x \ . \eeq
If we consider a change of variable which does not affect (but possibly depend on) both the $t$ and $w$ variables but only the $x$, i.e.
\beql{eq:cov} y \ = \ \psi (x,t;w) \ , \eeq
say with inverse $x = \xi (y,t;w)$, then we have
$\pa_x = ( \pa \psi / \pa x ) \pa_y$. It results immediately
\beq X \ = \ \vphi [ \xi (y,t;w),t;w] \, \( \frac{\pa \psi}{\pa x} \) \, \pa_y \ ; \eeq
if we choose $\psi$ such that
\beql{eq:phi1} \vphi [ \xi (y,t;w),t;w] \, \( \frac{\pa \psi}{\pa x} \) \ = \ 1 , \eeq
then the symmetry reads, in the new variables,
\beql{eq:Xsty} X \ = \ \pa_y \ . \eeq
The equation for $y$ will then be of the form
\beql{eq:KIto} d y \ = \ F (t;w) \, dt \ + \ S (t;w) \, dw \eeq
and hence is immediately integrable. The fact that $F$ and $S$ do not depend on $y$ is a direct consequence of having \eqref{eq:Xsty} as a symmetry, and this is turn depends on having \eqref{eq:Xst} as a symmetry of the original equation, \emph{and} on the fact that symmetries are conserved under a change of variables.\footnote{This is not immediately obvious, given that vector fields transform under the chain rule and equations transform under the Ito rule; but it can be proven, see \gcite{GL1,GL2}. Alternatively, it becomes obvious considering the Stratonovich representation of the SDE; but one should then prove that symmetries of Ito and Stratonovich equations coincide.}

For \eqref{eq:phi1} to be satisfied, we just choose
\beql{eq:kozmap} \psi(x,t;w) \ = \ \int \frac{1}{\vphi (x,t;w)} \ d x \ . \eeq
This, or more precisely the change of variable \eqref{eq:cov} with this choice for $\psi$, is the \emph{Kozlov  substitution} which integrates a SDE possessing a standard symmetry.

We note that \eqref{eq:phi1} can also be written in the form
$$ X (\psi ) \ = \ 1 \ ; $$ it goes without saying that we also have
$$ X(t) \ = \ 0 \ , \ \ X (w) \ = \ 0 \ . $$

From this point of view, the problem of integrating a symmetric SDE (using standard symmetries) reduces to a problem in the theory of characteristics for first order PDEs \gcite{ArnODE,ArnodeS}.

\medskip\noindent
{\bf Remark 1.} It should be noted that when $\vphi = \vphi (x,t)$, i.e. we have a \emph{deterministic standard symmetry}, the Kozlov substitution maps the original Ito equation into a new (integrable) Ito equation; whereas if $\vphi$ depends effectively on $w$, i.e. is \emph{random standard symmetry} the Kozlov substitution maps the original Ito equation into an integrable equation of generalized Ito type (see \gcite{GS17,Koz18a}; see also the discussion in \gcite{GR22a,GR22b,GR23a,GKS}). This means that when considering multiple symmetry reduction of a proper Ito equations, we are in general (that is, unless all the considered symmetries are not only standard ones, but also deterministic ones) led to consider generalized Ito equations and their symmetries. \EOR

\medskip\noindent
{\bf Remark 2.} It may be worth showing in detail that a symmetry of the form \eqref{eq:Ito} or \eqref{eq:gIto} implies that the equation is written -- after the Kozlov substitution -- as in \eqref{eq:KIto}. In fact, the determining equations for the Ito equation \eqref{eq:Ito}, or for the generalized Ito equation  \eqref{eq:gIto}
(we recall that, as just remarked, the transformation \eqref{eq:kozmap} could map the original Ito equation into a generalized one) are, in the scalar case we are considering,
\begin{eqnarray*}
& & \vphi_t \ + \ f \, \vphi_x \ - \ \vphi \, f_x \ + \ (1/2) \, \Delta (\vphi ) \ = \ 0 \ , \\
& & \vphi_w \ + \ \s \, \vphi_x \ - \ \vphi \, \s_x \ = \ 0 \ . \end{eqnarray*}
In the present framework, these should be seen as equations for $f = f(x,t)$ and $\s = \s (x,t)$ with $\vphi = 1$ given; note this implies $\Delta (\vphi) = 0$. Thus they just read
$$ f_x \ = \  0 \ , \ \ \ \ \s_x \ = \ 0 \ . $$
This implies that $f(x,t;w) = F(t;w)$, $\s (x,t;w) = S(t;w)$, as stated. \EOR

\medskip\noindent
{\bf Remark 3.} We can also easily compute the explicit form of the transformed equation. If $dx = f dt + \s dw$ and $y = \Psi (x,t,w)$, then
\begin{eqnarray*}
dy &=& \Psi_x \, dx \ + \ \Psi_t \, dt \ + \ \Psi_w \, dw \ + \ \frac12 \, \De (\Psi) \, dt \\
&=& \( \Psi_t \, + \, f \, \Psi_x \, + \, \frac12 \De (\Psi) \) \, dt \ + \ \( \Psi_w \, + \, \s \, \Psi_x \) \, d w \\
&:=& F \, dt \ + \ S \, dw \ . \end{eqnarray*}
This also shows that if $\Psi$ is chosen as in \eqref{eq:kozmap} and $\vphi$, and hence\footnote{Actually, \eqref{eq:kozmap} defines $\Psi$ up to a constant of integration, i.e. an arbitrary function of $w$ and $t$; we are taking this to be zero.} $\Psi$, does not depend on $w$ -- as happens for \emph{deterministic} standard symmetries -- then $F$ and $S$ will also be independent on $w$. On the other hand, if $\vphi$ depends on $w$, then in general $\Psi$ will also (whatever the integration constant one could introduce) depend on $w$, and so will $F$ and $S$. This shows that, as anticipated, in general the Kozlov substitution associated to a random standard symmetry will map a proper Ito equation for $x$ into a generalized Ito equation for $y$. \EOR

\medskip\noindent
{\bf Remark 4.} It is clear that the equations obtained under the Kozlov substitution for a deterministic standard symmetry are immediately integrated: in fact, we have
$$ y (t) \ = \ y(t_0) \ + \ \int_{t_0}^t F (\tau) \, d \tau \ + \ \int_{t_0}^t S (\tau) \, d w (\tau) \ , $$ were the first integral is a standard one and the second an Ito integral.

In the case of a random standard symmetry we also obtain a (formally) integrable, albeit non proper Ito, equation; the solution being
$$ y(t) \ = \ y(t_0) \ + \ \int_{t_0}^t F [\tau , w(\tau) ] \, d \tau \ + \ \int_{t_0}^t S [\tau , w(\tau)] \, d w (\tau) \ . \eqno{\odot} $$
\bigskip

The following two examples (and other ones later on) are taken from \gcite{GSW}.

\medskip\noindent
{\bf Example 1.} Consider the (proper) Ito equation
\beq dx \ = \ \[ e^{-x} \, - \, \frac12 \, e^{-2x} \] \, dt \ + \ e^{-x} \, dw \ . \eeq
It admits the deterministic standard symmetry
$$ X \ = \ e^{- x} \ \pa_x \ ; $$
upon passing to the variable
$$ y \ = \ \int e^x \ dx \ = \ e^x \ , $$
so that $X = \pa_y$, we get the new proper Ito equation
\beq d y \ = \ dt \ + \ dw \eeq which is readily integrated. \EOE

\medskip\noindent
{\bf Example 2.} The proper Ito equation
\beq dx \ = \ e^x \, dt \ + \ dw \eeq admits no deterministic standard symmetries, but admits the random standard symmetry
$$ X \ = \ e^{x-w} \, \pa_x \ . $$
With the Kozlov substitution, i.e. passing to the variable
$$ y \ = \ \int e^{w - x} \ dx \ = \ - \ e^{w-x} \ , $$
it is transformed into
\beq dy \ = \ e^w \ dt \ . \eeq
This is a generalized Ito equation, and it yields
$$ y(t) \ = \ y(t_0) \ + \ \int_{t_0}^t \exp[ w (\tau)] \ d \tau \ . $$
Note that the original equation is also (trivially) invariant under $\pa_t$ and $\pa_w$; these however are not admitted symmetries (see the discussion in \gcite{GSW}). In particular, $\pa_t$ expresses the fact the equation is time-autonomous and is thus trivial, while the translation generator $\pa_w$ maps the Wiener process $w(t)$ into a process with non-zero mean. \EOE

\subsection{W-symmetries}
\label{sec:strW}

We want now to consider W-symmetries, and attempt to proceed along the same lines.
In this case the vector fields generating the symmetries are of the form
\beql{eq:XW} X \ = \ \vphi (x,t;w) \, \pa_x \ + \ r \, w \, \pa_w \ . \eeq
We can apply the same type of reasoning as above; but now we should also allow for a change in the variable expressing the driving Wiener process. For greater generality we could also consider a change in the time variable (which would then become a random variable itself), i.e. consider the change of variables $(x,t;w) \to (y,u;z)$ with
\begin{eqnarray}
y &=& \psi (x,t;w) \ , \nonumber \\
u &=& \theta (x,t;w) \ , \label{eq:yuz} \\
z &=& \zeta (x,t;w) \ . \nonumber \end{eqnarray}
In the new variables, we have
\beq X \ = \ [ X(\psi)] \, \pa_y \ + \ [X(\theta)] \, \pa_t \ + \ [X (\zeta)] \, \pa_z \ . \eeq
Thus the straightening change of variables is determined solving the problem
\beql{eq:Wstraight} X(\psi) \ = \ 1 \ ; \ \ X(\theta) \ = \ 0 \ , \ \ X(\zeta ) \ = \ 0 \ . \eeq
It is immediately apparent that we can just choose $\theta (x,t;w) = t$, i.e. leave unchanged the time variable,  which thus retains its special role (for a discussion of the general case, see \gcite{GR22b}); but we cannot avoid to transform the $w$ variable.

In the case where $\vphi$ in \eqref{eq:XW} does not depend on $w$, which we will denote as \emph{split W-symmetries}\footnote{Since now the $(x,t)$ and the $w$ variables transform independently of each other under the action of $X$. Split W-symmetries were already introduced and considered in \gcite{GSW}; see there for some of their properties. A discussion of the use of (split) W-symmetries in the integration of proper Ito equations is also given in Appendix B to \gcite{GR22b}.}), the corresponding equations can be solved by the method of characteristics producing a general formula, i.e.
\begin{eqnarray}
\psi (x,t;w)  &=& \int \frac{1}{\vphi (x,t) } \ dx \ + \ \a ( \chi , t ) \ , \label{eq:psi} \\
\zeta (x,t;w) &=& \b (\chi , t ) \ . \label{eq:zeta} \end{eqnarray}
Here $\a$ and $\b$ are arbitrary functions (provided they define a non-singular change of variables, see below) of $t$ and of the characteristic function $\chi$, given by
\beql{eq:chi} \chi \ := \  w \ \exp\[ - \, \int \frac{r}{\vphi (x,t)} \, dx \] \ . \eeq

The requirement that these function $\psi$ and $\zeta$ define a proper change of variables rules out the possibility to choose $\b \equiv 0$, or more generally $\b = const$ (note that we can instead choose $\a \equiv 0$). Thus the simplest choice is
$$ \b (\chi,t) \ = \ \chi \ . $$
In any case, it follows from the previous general formulas that the new random variable $z = \zeta (x,t;w)$ is \emph{not} a Wiener process, and actually its statistical properties depend on the process $x(t)$ (that is, on the solution to our SDE) itself.

In other words, albeit we will in this way manage to write our equation in the form
\beq d y \ = \ F (t;z) \, dt \ + \ S(t;z) \, dz \ , \eeq
not only this equation will not be in proper Ito form (since the drift and the noise coefficient will depend on the driving process $z$), but moreover the driving process $z(t)$ will \emph{not} be a Wiener one, and its statistical properties are undetermined.

\medskip\noindent
{\bf Remark 5.} We would like to clarify one point. We have allowed a  reparametrization of the Wiener process $w(t)$; the form \eqref{eq:XW} of allowed generators clarifies that this corresponds to a \emph{rescaling} of $w(t)$. This poses a problem, as $\langle [w(t_1) - w(t_0)]^2 \rangle = |t_1 - t_0|$, while a rescaling by a factor $\la$ would introduce a factor $\la^2$ in this formula, i.e. transform $w(t)$ into a process which is not, strictly speaking, a Wiener one. On the other hand, it is also clear that a map $w \to \la w$ can be reinterpreted -- as far as the Ito equation \eqref{eq:Ito} is concerned -- as a map leaving $w$ untouched and mapping $\s$ into $\la \s$. It is for this reason that rescalings of $w(t)$ can be accepted. We refer to \gcite{GSW} for a more detailed discussion. \EOR

\medskip\noindent
{\bf Remark 6.} The argument used in Remark 5 becomes less simple in the case of \emph{generalized} Ito equations. In fact, in this case a rescaling of $w$ does on one hand produce a rescaling of $d w$, which can be discharged into a multiplying factor for the noise coefficient $\s$. But on the other hand such a rescaling will also, in general (i.e. unless the dependence of these on $w$ is very simple), affect the functional form of $f = f(x,t,w)$ and of $\s = \s (x,t,w)$. We cannot exclude \emph{apriori} that despite this the equation remains invariant -- through the change in the $x$ variable -- but we should expect W-symmetries to be even more rare for generalized Ito equations than for standard ones. \EOR

\medskip\noindent
{\bf Remark 7.}
We note that a different kind of full straightening can also be considered: that is, search for new variables $(y,z)$ -- with again $z$ the driving stochastic process and $y$ the driven dynamical random variable -- such that in these the vector field reads as $X = \pa_z$. This corresponds to just interchanging the role of $y$ and $z$ seen above, and by just the same computations we get that again the new driving process is in general \emph{not} a Wiener one. Thus the same considerations apply. \EOR

\medskip\noindent
{\bf Remark 8.} We have focused on \emph{split} W-symmetries; this assumption allowed to solve the characteristic equation and determine explicitly (through an elementary integration) the characteristic variable $\chi$. Needless to say, if we consider a non-split W-symmetry, the whole situation gets more involved, and determining the functions $\psi$ and $\zeta$ solving \eqref{eq:Wstraight} might be a highly nontrivial task. In any case, straightening will require introducing a new driving process $z(t)$ which will depend on both $x(t)$ and $w(t)$, and which will not be a Wiener process. So the observation that the transformed equation is not an Ito one in any sense remains valid. \EOR

\medskip\noindent
{\bf Remark 9.} As discussed above, we can not find a change of variables $(x,w) \to (y,z)$ taking a W-symmetry vector field $X = \vphi \pa_x + r w \pa_w$ to the form $X = \pa_y$ without ending up with a stochastic equation whose driving noise depends on the solution to the equation itself (which makes it untractable).

The situation is different if we want to take the vector field $X$ to a different standard form. In particular, if $X$ is a \emph{split W-symmetry} we can always take it in a harmless way to the form $X = \pa_y + r w \pa_w$ by the Kozlov substitution \eqref{eq:kozmap}; or to the scaling form $X = r (y \pa_y + w \pa_w)$ by the \emph{modified Kozlov substitution}
\beql{eq:mkozsub} y \ = \ \exp \[ \int \frac{r}{\vphi (x,t)} \ d x \] \ . \eeq This will be shown and discussed in Section \ref{sec:Wscaling}. Unfortunately  these forms, including the scaling standard form, seem to be useless as far as integrating the equation is concerned -- at least in the usual way.\footnote{It is conceivable that different way of integrating SDEs exist, in the same way as for deterministic equations (think of the Arnold-Liouville versus the Lax approach). In this sense, it can be hoped that (generalized) Ito equations which are invariant under a scale transformation can be handled more effectively than generic equations, e.g. using renormalization group techniques to look at the equilibrium distribution (see e.g. \gcite{Bir}), or a path integral approach (see e.g. \gcite{Cug}).
} \EOR

\medskip\noindent
{\bf Remark 10.}
One should also remember that integration (or reduction, in the case of system) is not the only use of symmetries; in particular, they also map solutions to solutions, and in this sense a W-symmetry shows that if we have solutions to a given equation for given realizations of the Wiener process, these will also provide solutions to the same equation for rescaled realizations of the Wiener process; or, more interestingly, solutions to a different equation (obtained by rescaling the noise coefficient $\s$; see Remark 5) for the same realization of the driving process. This will be discussed in Section \ref{sec:relW}. \EOR

\medskip\noindent
{\bf Example 3.} Consider the equation (with $a$ and $b$ both nonzero, but possibly constant)
\beql{eq:ex3} dx \ = \ \[ a (t) \ x \] \, dt \ + \ \[ b (t) \ \sqrt{x} \] \, dw \ ; \eeq
note that the half-line $R_+$ is invariant under this dynamics, so we can restrict to $x \ge 0$ (and dispense with absolute value signs in the argument of logarithms).

This equation admits symmetries with generator
\beq X \ = \ \vphi (x,t;w) \, \pa_x \ + \ r \, w \, \pa_w \eeq with $r$ an arbitrary number, provided we choose
\beq \vphi \ = \ 2 \, r \, x \ ; \eeq
thus we have W-symmetries, while no standard symmetries are present. We can, with no loss of generality, take $r=1$, and just consider
\beql{eq:X.ex3} X \ = \ 2 \, x \, \pa_x \ + \ w \, \pa_w \ . \eeq

The characteristic variables for this vector field are time $t$ and
\beql{eq:z.ex3} z \ := \ w/\sqrt{x} \ . \eeq
In order to straighten the vector field, we should pass to the variable
\beq y \ = \ \frac{1}{r} \ \log [ \sqrt{x}] \ ; \eeq
in fact, in these variables we have
$$ X \ = \ \pa_y \ . $$
Let us thus consider the change of variables
$ (x,t,w) \to (y,t, z)$.
We have, by Ito calculus,
\begin{eqnarray*}
dy &=& \( \frac{1}{4 x} \) \ \[ ( 2 a x \, - \, b^2) \, dt \ + \ 2 , b  \sqrt{x} \, dw \] \ , \\
dz &=& \( \frac{1}{8 x \sqrt{x} } \) \ \[ \(3 b^2 w \, - \,  4 a x w \, - \, 4 b \sqrt{x} \) \, dt
\ + \ \( 8 x \, - \, 4 b w \sqrt{x} \) \, d w \] \ . \end{eqnarray*}
We can use the second equation to express $dw$ in terms of $dz$ and $dt$; with this, and considering also the inverse change of variables
$$ x \ = \ e^{2 y} \ , \ \ \ w \ = \ e^y \ z \ , $$
we obtain the transformed equation, which reads
\beql{eq:y.ex3} d y \ = \ \( \frac{8 \, a \ - \ b \, e^{- 2 y} \, z}{8 \ (2 \ - \ b \, z)} \) \, d t \ + \ \( \frac{b}{2 \ - \ b \, z } \) \, d z \ . \eeq
In this equation, the drift and noise coefficients are both independent of $y$; thus straightening the vector field led in fact to an equation which is \emph{formally} integrable. Note that both the drift and the noise coefficients depend on $z$. Moreover, and more relevantly, as obvious from the explicit expression \eqref{eq:z.ex3} of $z$, this is \emph{not} a Wiener process. Hence \eqref{eq:y.ex3} is \emph{not} an Ito equation, not even in the generalized sense.\footnote{The formulas above are slightly simplified by taking $ a (t) = b (t) = 1$; this simplified setting does not change the relevant facts.}

Thus it makes no sense to wonder if the transformed vector field $X = \pa_y$ formally satisfies the determining equations for symmetries of the transformed equation (which in reality is not the case), as we do not know how to compute symmetries of a non-Ito equation. \EOE

\medskip\noindent
{\bf Example 4.} Consider again eq. \eqref{eq:ex3} and its symmetry \eqref{eq:X.ex3}. Passing to variables
\beq y \ = \ \frac{w}{\sqrt{x}} \ , \ \ \ z \ = \ \frac{1}{r} \ \log \[ \sqrt{x} \] \ , \eeq
this reads $X = \pa_z$. In fact, via standard Ito calculus, the equation reads now
\beql{eq:dy.ex4} d y \ = \ \frac{b}{8} \, e^{- z} \, y \ dt \ . \eeq
This looks again like a generalized Ito equation, but again one should note that $z$ is \emph{not} a Wiener process and thus \eqref{eq:dy.ex4} is \emph{not} an Ito equation in any sense. \EOE

\subsection{Discussion}
\label{sec:discstraight}

Summarizing our discussion in this Section, we have shown that:

\begin{itemize}

\item Random standard symmetries of a proper Ito equation determine, through the Kozlov substitution \eqref{eq:kozmap} $x \to y$, a map to a generalized Ito equation for the random dynamical variable $y(t)$ in which both the drift and noise coefficients do not depend on $y$, and which is therefore (at least formally) integrable; considering the change of variables inverse to Kozlov one, i.e. $y \to x$, we obtain a solution to the original equation. In this sense random standard symmetries are (nearly) ``as good as deterministic standard symmetries'' for what concerns integration of an Ito equation.

\item W-symmetries produce, through straightening of the symmetry vector field via the Kozlov substitution \eqref{eq:kozmap}, a map to an equation which is in no sense an Ito equation. This entails that W-symmetries are of no use in the integration (or reduction, in the case of systems) of Ito equations through passing to symmetry-adapted variables.\footnote{One cannot exclude they can be used for integration of the equation via a different mechanism, see also the footnote to Remark 9. I am not aware of any result in this direction.}

\end{itemize}

\section{W-symmetries in standard scaling form}
\label{sec:Wscaling}

In Section \ref{sec:straight} we have considered straightening of the symmetry vector field in the same sense as in the straightening theorem (or flow box theorem) in Dynamical Systems. That is, we searched for a change of variables transforming the flow of the symmetry vector field in a rigid translation (in one of the variables, i.e. the dynamical variable).

We can modify this approach and look for a change of variables mapping the symmetry vector field into one with a definite (and convenient) form, not necessarily a translation. In particular, we can consider changes to variables in which the symmetry vector field is a \emph{scaling} one\footnote{We note in passing that this is also, usually, a preliminary step in the proof of the straightening (or ``flow-box'') theorem \gcite{ArnODE}.}.

It turns out this problem can be solved in a simple way, as described below; unfortunately, the solution does not give any help towards integration of an equation possessing a W-symmetry.

We look for functions $\psi$ and $\zeta$, see \eqref{eq:yuz}, such that in the new variables $(y,t,z)$ we have
\beql{eq:XS} X \ = \ r \, y \, \pa_y \ + \ r \, w \, \pa_w \ . \eeq
These are obtained as solutions to the set of equations
\beq X (\psi) \ = \ r \, \psi \ , \ \ \ X(t) \ = \ 0 \ , \ \ \ X (\zeta ) \ = \ r \, \zeta \ . \eeq
It should be noted that the last two equations are satisfied if we leave $z = \zeta (x,t,w) = w$, so that we only have to look for a change in the dynamical variable, which changes from $x$ to $y = \psi (,t,w)$. This guarantees automatically that we do not affect the Wiener character of the driving process, and hence the major problem encountered above (see Sect.\ref{sec:strW}) will not show up.

Thus we only have to solve $X (\psi ) = r \psi$. Here again \emph{split W-symmetries} lead to a simpler problem. In fact, for
\beql{eq:Xsplit} X \ = \ \vphi (x,t) \, \pa_x \ + \ r \, w \, \pa_w \eeq the solution to $X (\psi ) = r \psi $ is given by \beq \psi (x,t,w) \ = \ \exp \[ r \ \int \frac{1}{\vphi (x,t) } \ dx \] \ + \ \alpha (\chi,t ) \ , \eeq
where the characteristic variable $\chi$ is as in \eqref{eq:chi}. We can choose $\a \equiv 0$, so that -- as in Remark 9, see \eqref{eq:mkozsub} -- we get
\beql{eq:mkozmap} y \ = \ \exp \[ r \, \int \frac{1}{\vphi (x,t)} \ dx \] \ := \ \psi (x,t) \ . \eeq
We will refer to the above as the \emph{modified Kozlov substitution}.
This substitution allows to get the symmetry vector field in the standard scaling  form \eqref{eq:XS}.

\medskip\noindent
{\bf Remark 11.} We can reverse our point of view and wonder what is the most general form of equations admitting \eqref{eq:XS} as a symmetry. This requires to solve the determining equations \eqref{eq:gendeteq1}, \eqref{eq:gendeteq2}  as equations for $f$ and $\s$ with $\vphi = r x$. This is an easy task, and we get
\beq f (x,t,w) \ = \ x \ \^f (\chi , t) \ , \ \ \s (x,t,w) \ = \ \^\s (\chi , t) \ ; \ \ \ \chi \ := \ w/x \ . \eeq
Note that if we want a \emph{proper} Ito equation, then necessarily $\^f (\chi , t) = \wt{f} (t)$ and $\^\s (\chi , t ) = \wt{\s} (t)$. This corresponds to a trivially integrable equation. The same holds \emph{a fortiori} if we just consider time-autonomous equations, in which case the equations we get are simply
$$ d x \ = \ \la \, x \, dt \ + \ \mu \, dw \ , $$ with $\la$ and $\mu$ real constants. \EOR

\medskip\noindent
{\bf Example 5.} Consider the equation
\beql{eq:ex5} dx \ = \ \la \, x \, dt \ + \ \mu \,  dw \ , \eeq
with $\la$ and $\mu$ two nonzero constants.
This admits symmetries of the form \eqref{eq:Xsplit} with
$$ \vphi (x,t,w) \ = \ r \, x \ + \ \kappa \, e^{\la t} \ , $$ where $\kappa$ is a constant.

It may be noted that setting $R=0$ we get a standard symmetry; if we use this (say with $\kappa = 1$)  we can -- as well known \gcite{Koz1,Koz2} -- integrate equation \eqref{eq:ex5} by using symmetry. In fact with the (standard) Kozlov substitution we get
\beql{eq:ex5.cov}  y \ = \ \int \frac{1}{\vphi} \ dx \ = \ \int e^{- \la t} \ dx \ = \ e^{- \la t} \ x \ ; \eeq correspondingly, \eqref{eq:ex5} is transformed into
\beql{eq:ex5.neweq} dy \ = \ \mu \ e^{- \la t} \ d w \ . \eeq
This is readily integrated,
$$ y(t) \ = \ y (t_0) \ + \ \mu \ \int_{t_0}^t e^{- \la \tau} \ d w (\tau) \ , $$ with $y(t_0) = e^{- \la t_0} x(t_0)$, and the solution to \eqref{eq:ex5} is provided by
\beq x(t) \ = \ e^{\la t} \ y (t) \ = \ x (t_0) \ + \ \mu \ e^{\la t} \ \int_{t_0}^t e^{- \la \tau} \ d w (\tau) \ . \eeq
The case where $\la$ and $\mu$ are not constants, but only depend on $t$, is dealt with in the same way.

Let us now focus on proper W-symmetries; we thus set $\kappa =0$, disregarding the standard part of the symmetry. We are left with
$$ X \ = \ r \ \( x \, \pa_x \ + \ w \, \pa_w \) \ . $$
In this case, the symmetry vector field is already in scaling form. Note also  that  applying the (modified) Kozlov substitution \eqref{eq:mkozmap} we just get
$$ y \ = \ \exp \[ \int \frac{r}{\vphi} \ dx \] \ = \ \exp \[ \log (x) \] \ = \ x  \ . $$

It is also interesting, in particular in connection with the discussion in Section \ref{sec:persW} to see how the symmetry vector fields are transformed under the change of variables. We have, as seen above, two symmetry generators; in the original coordinates these are written as
$$ X_0 \ = \ e^{\la t} \, \pa_x \ , \ \ \ X_1 \ = \ x \, \pa_x \ + \ w \, \pa_w \ ; $$ they satisfy the commutation relation
$[ X_0 , X_1 ] = X_0 $.
With the change of variable \eqref{eq:ex5.cov}, we get
$$ X_0 \ = \ \pa_y \ , \ \ \ \ X_1 \ = \ y \, \pa_y \ + \ w \, \pa_w \ . $$
It is immediate to check that the transformed equation \eqref{eq:ex5.neweq} still admits both $X_0$ and $X_1$ as symmetry generators. \EOE

\section{Persistence of W-symmetries under changes of variables}
\label{sec:persW}

As mentioned in the Introduction and in Section \ref{sec:straight}, we are not sure \emph{apriori} that symmetries of an equation, including W-symmetries, remain such after a change of variables, both in general and when considering specifically the Kozlov substitution \eqref{eq:kozmap}, or the modified Kozlov substitution \eqref{eq:mkozmap}.

This question can be investigated using the \emph{Stratonovich equation} \gcite{\sderef} corresponding to the Ito equation we are considering, i.e. \eqref{eq:Strat}. Thus the discussion in this Section is strongly related to that of the following Section \ref{sec:relation}.

We note that -- as well known and already mentioned -- a Stratonovich equation transforms geometrically, i.e. following the familiar chain rule, under diffeomorphisms; in fact this is the main advantage of the Stratonovich formulation (modulo the usual warning about subtleties arising when one considers in more detail the correspondence between Ito and Stratonovich equations \gcite{Stroock}). Thus Stratonovich equation and symmetry vector fields transform in the same way, and it follows that symmetries of a Stratonovich equation are obviously persistent under diffeomorphisms.

In this way our question can be reformulated as the problem of determining if W-symmetries of an Ito equation are also W-symmetries of the corresponding Stratonovich equation; if this is the case, we are guaranteed that symmetries of the Ito equation are preserved under smooth changes of variables. We recall that the problem has been solved in general for deterministic standard symmetries in the Thesis by C.Lunini (unpublished); see also \gcite{GL1,GL2}. The proof for random standard symmetries can be obtained along the same lines as the one for deterministic ones, and is given in Sect.\ref{sec:relation}.

As for W-symmetries, it was observed already in \gcite{GSW} that in general this is \emph{not} the case. In particular, Theorem 1 and its Corollary in \gcite{GSW} dealt with the $n$-dimensional case, hence with the Ito systems
\beql{eq:ItoNdim} d x^i \ = \ f^i (\xb , t) \, dt \ + \ \s^i_{\ j} (\xb, t) \, d w^j \ ; \eeq they tell us that we should look more carefully at the R matrix embodying the coefficients appearing in the map $w^i \to z^i = R^i_{\ j} w^j$ associated to the action on $w$ variables of a W-symmetry vector field
$$ X \ = \ \vphi^i (\xb , y , \wb ) \, \frac{\pa}{\pa x^i} \ + \ R^i_{\ j} w^j \, \frac{\pa}{\pa w^i} \ ; $$ note that in the one-dimensional case only dilation type W-symmetries can be present.

\medskip\noindent
{\bf Proposition} (Theorem 1 in \gcite{GSW}). {\it All the rotation linear W-symmetries of an Ito equation are also symmetries of the associated Stratonovich equation, and vice
versa. Dilation W-symmetries of an Ito equation are also symmetries of the associated Stratonovich equation (and vice versa) if and only if the
diffusion matrix is spatially constant.}

\medskip\noindent
{\bf Corollary} (Corollary 1 in \gcite{GSW}). {\it If the diffusion matrix $S$ with entries $\s^i_k$ in \eqref{eq:ItoNdim} is constant with respect to space variables, then all W-symmetries of the Ito equation are also
symmetries of the corresponding Stratonovich equation.}
\bigskip

These results can be restated in the present restricted scalar case -- thus dealing with \eqref{eq:Ito} rather than with the system \eqref{eq:ItoNdim} -- as follows (note that for $\s_x = 0$ we have $b = f$):

\medskip\noindent
{\bf Lemma 1.} {\it W-symmetries of the proper Ito equation \eqref{eq:Ito} are also W-symmetries of the associated Stratonovich equation \eqref{eq:Strat} if and only if $\s_x = 0$.}

\medskip\noindent
{\bf Corollary.} {\it For $\s_x = 0$, W-symmetries of the Ito equation \eqref{eq:Ito} are preserved under diffeomorphisms. This is in general not the case if $\s_x \not= 0$.}
\bigskip

We will give a fresh discussion of this point, i.e. the interrelations between symmetries of an Ito equation and of the associated Stratonovich equation in the scalar case, in the next Section \ref{sec:relation}; this will also consider generalized Ito and Stratonovich equations. For the moment, it suffices to note that our Corollary means we are not guaranteed symmetries determined in the analysis of an Ito equation -- or a system of Ito equations -- will be still present after we pass to symmetry adapted variables (both when these are adapted to a different symmetries, and also when they are adapted to the very symmetry we are looking at). This point will be made clear by Example 6.

\medskip\noindent
{\bf Remark 11.} Our conclusion in this respect is that even if one was ever able to find a way to integrate Ito equations through W-symmetries, they could be safely used to this aim \emph{only} in the case where the equations under study  have spatially constant noise coefficient. \EOR

\medskip\noindent
{\bf Example 6.} We consider a slight -- but, as we will see, not innocent -- generalization of Example 5 above, i.e. the equation \beql{eq:ex6} dx \ = \ \la \, x \, dt \ + \ \mu \, x^\a \, dw \ , \eeq
with $\a \not= 0$ a new constant (for $\a = 0$ we get Example 5 again). We will assume also $\a \not= 1$, for a reason arising in a moment.

This equation admits the W-symmetries generated by
\beql{eq:ex6X} X \ = \ r \ \( x  \, \pa_x \ + \ (1 \, - \, \a ) \, w \, \pa_w \)  \ ; \eeq
such a vector field generates the one-parameter scaling group ($\ga \in {\bf R}$) \beql{eq:ex6scal} x \ \to \ \ga \, x \ , \ \ \ w \ \to \ \ga^{1 - \a} \, w \ . \eeq Note that for $\a = 1$ we are reduced to a standard symmetry, actually a deterministic one; we will hence assume $\a \not= 1$. Note also that for $\a = 0$ the vector field $X$ is already in standard scaling form.

We will now consider the change of variable taking $X$ to its standard scaling form
\beql{eq:ex6XX} X \ = \ r \ \( y \, \pa_y \ + \ w \, \pa_w \) \ , \eeq and enquire if it will still be a symmetry of the transformed equation. Note that to fit our general formalism for the modified Kozlov substitution we should rewrite the vector field \eqref{eq:ex6X} in the equivalent form
\beq X \ = \ \frac{r}{1-\a} \ x \ \pa_x \ + \ r \ w \ \pa_w \ , \eeq i.e. we have
\beq \vphi \ = \ \( \frac{r}{1 - \a} \) \ x \ . \eeq

The modified Kozlov substitution \eqref{eq:mkozmap} yields then
\beq y \ = \ \exp \[ \int \frac{r}{\vphi} \ dx \] \ = \ x^{(1 - \a )} \ ; \eeq
as expected, for $\a=0$ we get $y=x$, while for $\a=1$ the result is not acceptable (this corresponds to the fact that for $\a=1$ we have no W-symmetry).
Ito calculus yields the transformed equation, i.e. the equation for the new variable, in the form
\begin{eqnarray}
dy &=& F \, dt \ + \ S \, dw \ ; \label{eq:ex6y} \\
F &:=& \frac{1-\a}{2} \ \[ 2 \, \la \, y \ - \ \a \, \mu^2 \, y^{- 1} \] \ , \nonumber \\
S &:=& (1 \ - \ \a) \ \mu \ . \nonumber \end{eqnarray}

When we express the determining equations -- and the vector field -- in terms of the new variables, we obtain that the second one, eq.\eqref{eq:gendeteq2}, is satisfied; the first one, \eqref{eq:gendeteq1}, reads instead
\beq \( - \, \frac{1}{y} \) \ (1 \ - \ \a ) \ \a \, \mu^2 \, r \ = \ 0 \ . \eeq

That is, the equation is satisfied, and hence our equation \eqref{eq:ex6} recast in the $(y,t,w)$ variables -- that is, in the form \eqref{eq:ex6y} -- admits the scaling vector field  \eqref{eq:ex6XX} as a symmetry only for $\a = 0$ (in this case, as we have seen, $y=x$ and indeed we just get $dy = \la y dt + \mu dw$) and for $\a = 0$; that is, respectively, only when we fall back into Example 5 or when $X$ is already in standard scaling form and does not act on $w$, see \eqref{eq:ex6X}, so that no change of variable should occur. \EOE

\medskip\noindent
{\bf Example 7.}
Consider the equation
\beql{eq:exx7.eq}  dx \ = \ \la \, dt \ + \ \mu \, dw \ , \eeq
with $\la$ and $\mu$ nonzero real constants (the cases with vanishing $\la$ or $\mu$ are trivial). As shown in Example 11 in \gcite{GSW}, this admits the W-symmetries
\beql{eq:exx7.X} X_\Theta \ := \ r \ \[ \( (x \, - \, \la \, t ) \ + \ \Theta (\zeta) \) \, \pa_x \ + \ w \, \pa_w \] \eeq where $\Theta$ is an arbitrary smooth function of its argument
$$ \zeta \ := \ w \ - \ \(\frac{1}{\mu}\) \, x \ + \ \( \frac{\la}{\mu} \) \, t \ . $$
If $\Theta \not= 0$, this is a non-split W-symmetry. One can check that for $\Theta \not= 0$ the modified Kozlov substitution \eqref{eq:mkozmap} (which, we recall, only applies for split W-symmetries) does \emph{not} provide a scaling form for $X_\Theta$.

On the other hand, choosing $\Theta = 0$ we get a split W-symmetry, and \eqref{eq:mkozmap}, which now reduces to
\beql{eq:exx7.y} y \ = \ x \ - \ \la \, t \ , \eeq
yields in fact the scaling form \eqref{eq:XS} for our $X$.

The transformed equation is just
\beql{eq:exx7.dy} dy \ = \ \mu \, dw \ ; \eeq
this is trivially invariant under $X = y \pa_y + w \pa_w$.
\EOE

\section{On the relation between determining equations for symmetries of generalized Ito versus Stratonovich equations}
\label{sec:relation}

We want now to discuss the relation between the determining equations for symmetries, in general W-symmetries, of a (generalized) Ito equation and those for the corresponding (generalized) Stratonovich equation. We will limit to consider, as always in the present paper, the one-dimensional case.

As already observed, the second equation in both sets, i.e. \eqref{eq:gendeteq2} and \eqref{eq:deteqstrat2}, is just the same.
This can be used to express $\vphi_w$ and its derivatives $\vphi_{xw}$, $\vphi_{ww}$ in terms of $\vphi$, $\vphi_x$ and $\vphi_{xx}$. In this way we get
\begin{eqnarray}
\vphi_w &=& - \phi_x \, \s \ + \ r \, \s \ + \ \vphi \, \s_x \ + \ r \, \s_w \, w  \ , \nonumber \\
\vphi_{xw} &=& - \vphi_{xx} \, \s \ - \ \vphi_x \, \s_x \ + \ r \, \s_x \ + \ \vphi_x \, \s_x \ + \ \vphi \, \s_{xx} \ + \ r \, \s_{xw} \,  w \ , \label{eq:phiwetc} \\
\vphi_{ww} &=& - \vphi_{xw} \,  \s \ - \ \vphi_x \, \s_w \ + \ r \, \s_w \ + \ \vphi_w \, \s_x \ + \ \vphi \, \s_{xw} \ + \ r \, \s_{ww} \, w \ + \ r \, \s_w \ . \nonumber \end{eqnarray}

We can then substitute these in the first equation \eqref{eq:gendeteq1}, using the explicit expression of the Ito Laplacian. In this way \eqref{eq:gendeteq1} reads, up to an overall $1/2$ factor,
\begin{eqnarray} & &  2 \phi_t - 2 f_x \vphi  + 2 f \vphi_x - \vphi_x \s_w + 2 r \s_w - \vphi_x \s \s_x + 2 r \s \s_x + \vphi \s_x^2  \nonumber \\
 & &  \ + \vphi \s_{xw} + \vphi \s \s_{xx}  - 2 f_w r w + r \s_{ww} w + r \s_w \s_x w + r \s \s_{xw} w  \ = \ 0 \ . \label{eq:isa} \end{eqnarray}

We can now look at the first equation in the Stratonovich case, i.e. \eqref{eq:deteqstrat1}. Here the Ito Laplacian plays no role, but we should substitute for $b$ and its derivatives by choosing the $b$ corresponding to $f$ for the Ito equation, i.e. according to the appropriate generalization of the Stratonovich map formula (see Appendix \ref{app:relationseqs} for its derivation), i.e. \begin{eqnarray}
b &=& f \ - \ (1/2) \, (\s \, \s_x \ + \ \s_w ) \ , \label{eq:stratform} \\
b_x &=& f_x \ - \ (1/2) \, \( \s \, \s_{xx} \ + \ \s_x^2 \ + \ \s_{xw} \) \ , \nonumber \\
b_w &=& f_w \ - \ (1/2) \, \( \s_w \, \s_x \ + \ \s \, \s_{xw} \ + \ \s_{ww} \) \ . \nonumber \end{eqnarray}
In this way \eqref{eq:deteqstrat1} reads (again up to an overall $1/2$ factor)
\begin{eqnarray}  & & 2 \vphi_t - 2 f_x \vphi  + 2 f \vphi_x - \vphi_x \s_w - \vphi_x \s \s_x + \vphi \s_x^2  \nonumber \\
& &  \ + \vphi \s_{xw} + \vphi \s \s_{xx} - 2 f_w r w + r \s_{ww} w + r \s_w \s_x w + r \s \s_{xw}  w  \ = \ 0 \ . \label{eq:isb} \end{eqnarray}

Now we are wondering if \eqref{eq:isb} is the same as \eqref{eq:isa}; or more precisely if they admit the same set of solutions. Subtracting \eqref{eq:isb} from \eqref{eq:isa}, or equivalently determining say $\vphi_t$ by \eqref{eq:isb} and substituting for it in \eqref{eq:isa}, we obtain the (simple, but nonlinear) condition
\beql{eq:isdiff} r \ \( \s_w \ + \ \s \, \s_x \) \ = \ 0 \ . \eeq
Several special cases are worth discussing.

\begin{enumerate}
\item First of all we note that $f_w$ does not appear here (at difference with $\s_w$); thus the difference between proper and generalized equations is actually present only depending on the $w$ dependence of the noise coefficient $\s$.
\item For standard symmetries (so $r=0$) the condition is identically satisfied. This confirms the results in \gcite{GS17,GSW} and \gcite{GL1,GL2}. Note that actually this also applies for generalized equations.
\item Consider proper equations ($f_w = 0 = \s_w$) and proper W-symmetries ($r \not= 0$). In this case our condition \eqref{eq:isdiff} reduces to
    $\s \s_x  = 0$. Thus it is satisfied if and only if $\s_x = 0$, i.e. if and only if the noise coefficient is spatially constant. This confirms the result in \gcite{GSW} (see Theorem 1 therein); see also \gcite{GR22a} (Appendix D therein). Note this is also the case if we have a generalized equation with $f_w \not= 0$ and $\s_w = 0$.
\item Finally, in the general case (generalized equations, proper W-symmetries) symmetries for the generalized Ito equations are also symmetries for the corresponding generalized Stratonovich equation if and only if the condition \eqref{eq:isdiff} is satisfied. Note that in this case, as we assumed $r \not= 0$, eq. \eqref{eq:stratform} yields $b = f$.
\end{enumerate}

\medskip\noindent
{\bf Remark 12.}
The condition \eqref{eq:isdiff} for proper W-symmetries -- that is, for $r \not= 0$ -- is just the Euler transport equation $u_t + u u_x = 0$ with $\s$ taking the place of the velocity field $u$ and $w$ taking the place of $t$ (in our case $t$ has a parametric role). As well known, it is not possible to obtain the general solution $u = u(x,t)$ for the Euler equation in explicit form; hence we are not able to provide the most general form of noise coefficients $\s (x,t,w)$ satisfying \eqref{eq:isdiff}. \EOR

\medskip\noindent
{\bf Remark 13.} It is well known (see e.g,. \gcite{Oksendal}) that a proper Ito equation can always be transformed into one with a unit noise coefficient, just by introducing the new dynamical variable
$$ \xi \ = \ \int \frac{1}{\s (x,t)} \, d x \ . $$ A simple computation shows that this is not possible for generalized Ito equations. Thus the $w$ dependence of the noise coefficient cannot be eliminated by this standard procedure.

One could of course attempt to generalize this by considering
$$ \xi \ = \ \int \Phi \, dx $$ with $\Phi$ to be determined depending on $\s$. However, it turns out the equation for $\xi$ will be of the form
$$ d \xi \ = \ F (x,t,w) \, dt \ + \ S (x,t,w) \, dw $$ with $F$ a function whose explicit expression is not relevant here, and
$$ S \ = \ \Phi \, \s \ - \ \int \Phi_w \, dx \ . $$ Thus in order to get $S=1$  we should solve an integro-differential equation for $\Phi$. Differentiating this w.r.t. $x$ we get
$$ \Phi_w \ = \ \pa_x \( \Phi \, \s \) \ , $$ but substituting back into our equation we get a trivial identity. \EOR

\medskip\noindent
{\bf Remark 14.} In the one-dimensional case, proper W-symmetries are quite rare for proper Ito equations, as discussed in \gcite{GR22a}, except for nearly trivial cases as the one considered in following Example 8 below.

On the other hand, one should bear in mind that we are requiring our vector field, in particular those generating W-symmetries, to leave the $t$ variable untouched. This may be too strict a requirement for W-symmetries. E.g., it is immediate to check that the equation \eqref{eq:exx7.eq} seen in Example 7 does not admit W-symmetries with this requirement; but it is equally immediate that dropping such a requirement we get the W-symmetry with generator
$$ X \ = \ x \, \pa_x \ + \ t \, \pa_t \ + \ w \, \pa_w \ . $$
This yields the one-parameter group
$$ x \to \la x \ , \ \ t \to \la \, t \ , \ \ w \to \la \, w \ ; $$
and it is immediate to check this leaves indeed \eqref{eq:exx7.eq} invariant. \EOR

\medskip\noindent
{\bf Example 8.} The proper Ito equation considered in Example 5
$$ d x \ = \ \la \, x \, dt \ + \ \mu \, dw \ , $$
with $\la$ and $\mu$ nonzero real constants, admits the proper W-symmetry with generator
$$ X \ = \ r \ \( x \, \pa_x \ + \ w \, \pa_w \) \ ; $$ this corresponds to the scaling group $(x,w) \to (s x , s w )$. It is immediate to see that $X$ is also a symmetry for the corresponding Stratonovich equation. Note in this case we have $\s_x = 0$. \EOE

\medskip\noindent
{\bf Example 9.} Consider again the proper Ito equation \eqref{eq:ex6} studied in Example 6, i.e. $d x =  \la x dt + \mu x^\a dw$, where we assume $\a \not= 0$, $\a \not= 1$. As seen in Example 6, this admits the W-symmetry \eqref{eq:ex6X}, which generates the one-parameter scaling group \eqref{eq:ex6scal}.

The Stratonovich equation associated to \eqref{eq:ex6} is
\beql{eq:ex6strat} dx \ = \ \( \la \, x \ - \ \frac12 \, \a \, \mu^2 \, x^{(2 \a - 1)} \) \, dt \ + \ \mu \, x^\a \circ dw \ . \eeq
It is immediately apparent that the scaling \eqref{eq:ex6scal} leaves invariant the Stratonovich equation \emph{only} for $\a = 1$, i.e. when our W-symmetry \eqref{eq:ex6X} reduces to a (deterministic) standard symmetry.

We can verify this fact also through the determining equations for symmetries of the Stratonovich equation, see \eqref{eq:deteqstrat1}, \eqref{eq:deteqstrat2} above. While \eqref{eq:deteqstrat2} is immediately satisfied (as obvious, since it is just the same as the corresponding determining equation for the Ito equation), the first one \eqref{eq:deteqstrat1} yields
\beq \a \, \( \a \ - \ 1 \) \ \mu^2 \ x^{2 \a - 1} \ = \ 0 \ , \eeq
which again is satisfied (recalling we need $\mu \not= 0$ in order to have a stochastic equation, and $\a \not= 0$ in order not to fall back into Example 5) only in the uninteresting case $\a = 1$, see above.

That is, in this case W-symmetries \eqref{eq:ex6X} of the Ito equations \eqref{eq:ex6} are \emph{not} symmetries of the associated Stratonovich equation. \EOE

\medskip\noindent
{\bf Example 10.}
Consider, as in \gcite{GSW}, the family of generalized Ito equations
\beql{eq:ex5bI} dx \ = \ \[ x \ + \ g (t,\zeta ) \] \ dt \ + \ \[ \frac{x \ + \ e^{-t} \ \varrho (t,\zeta)}{w} \] \ dw \ , \eeq
where $g$ and $\varrho$ are arbitrary smooth functions of their arguments and we have defined
$$ \zeta \ := \ w \ - \ e^{- t} \, x \ . $$
In this case the symmetry determining equations \eqref{eq:gendeteq1}, \eqref{eq:gendeteq2} admit the solution $\vphi = r e^t w$, i.e. the proper (for $r \not= 0$) W-symmetry generated by
$$ X \ = \ r \ \( e^t \, w \, \pa_x \ + \ w \, \pa_w \) \ ; $$
note that $X(\zeta) = 0$.

When considering the corresponding generalized Stratonovich equation we get, by \eqref{eq:stratform},
$$ b(x,t,w) \ = \ [x \ + \ g(t,\zeta )] \ - \ \frac{e^{-t}}{2 \, w} \ \varrho_\zeta (t,\zeta) \ + \ \frac{1}{2 \, w^2} \ \[e^{- 2 t} \, x \, \rho_\zeta (t,\zeta) \ + \ e^{- 3 t} \, \varrho (t,\zeta) \, \varrho_\zeta (t,\zeta) \]  \ . $$ Plugging this into the first determining equation \eqref{eq:deteqstrat1} for the Stratonovich equation corresponding to \eqref{eq:ex5bI}, we get
$$ \( \frac{r \ e^{-t}}{w^2} \) \ \[ \( e^{2 t} \, \zeta \ - \ \varrho (t,\zeta) \) \ \rho_\zeta (t,\zeta) \] \ = \ 0 \ . $$
As we assume $r \not= 0$, this is satisfied only for
$$ \varrho (t,\zeta ) \ = \ \begin{cases} \varrho^{(a)} (t,\zeta) \ = \ \eta (t) &, \\
\varrho^{(b)} (t,\zeta) \ = \ e^{2 t} \ \zeta & . \end{cases}   $$
In these cases we get respectively
$$ \s (x,t,w) \ = \  \begin{cases} \s^{(a)}  \ = \ [x \ + \ e^{- t} \, \eta (t)]/w &, \\
\s^{(b)} (t,\zeta) \ = \ e^{t} & . \end{cases}  $$
It is immediate to check that \eqref{eq:isdiff} is satisfied in both cases. \EOE

\section{On the relevance of W-symmetries}
\label{sec:relW}

The net result of our study, for what concerns W-symmetries, is that they are no good at integrating Ito equations. Strictly speaking, we have been discussing only \emph{scalar} equations, but it is clear that the problem met in this setting will also appear for systems of equations.

This should not be a surprise: in the scalar case W-symmetries can act on the driving Wiener process only by a scaling. But already when W-symmetries were introduced \gcite{GSW}, and the limitations on the possible form of the $R$ matrix determined, it was clear that this, being a matrix in the algebra of the linear conformal group, is made of a combination of a rotation matrix and of a dilation one. And it was observed there (see Section VII.C therein, in particular theorems 1 and 2) that while rotation type W-symmetries behave nicely under change of variables, and are common to a proper Ito equation and the associated Stratonovich one, the same does not hold for dilation type W-symmetries.

Thus our present discussion confirms the shortcomings of W-symmetries when it comes to scalar equations, and in general to dilation type W-symmetries; note that while in previous work \gcite{GSW} only proper Ito equations were considered, here we dealt also with generalized ones.

On the other hand, one should bear in mind that this lack of effectiveness of W-symmetries in the scalar case is related, indeed, to the fact they are ne\-ces\-sa\-ri\-ly of dilation type; in other words, in higher dimension there is space for the (rotation type) W-symmetries to come handy.

In fact, it suffices to consider the stochastic isotropic linear oscillator (here $\kappa$ and $\s$ are positive real constants)
\beq dx^i \ = \ - \kappa \, x^i \, dt \ + \ \s \, d w^i \eeq
discussed at length in \gcite{GSW} (see in particular Example 7 therein). If we are not considering W-symmetries, the (obvious) rotational symmetry of the system can not be taken into account.

But, there are other ways in which W-symmetries may be useful. In fact, one should recall that integration of a symmetric equation is only one of the ways in which symmetries can be used. Another way is that symmetries map solutions into solutions, so once we have determined  a solution to a symmetric equation, we immediately obtain also all the symmetry-related solutions. This is a well known fact for deterministic equations, and also for standard symmetries of stochastic equations; but in the case of stochastic equations and proper W-symmetries, it is worth discussing it in some detail, both for proper and generalized Ito (or Stratonovich) equations.

First of all, Remark 5 shows that a proper W-symmetry should actually be seen as mapping a proper Ito equation $dx = f (x,t) dt + \s (x,t) dw$ into a \emph{different} Ito equation, with a rescaled noise coefficient $\s$ and the same driving process. This shows that in this case a W-symmetry will map a solution to the original equation for a given realization of the driving process into a solution to the \emph{transformed} equation for the \emph{same} realization of the driving process. This has only trivial applications in the scalar case (see Examples 11 and 12 below), but in higher dimension this is essential to recover the vectorial character of isotropic Ito equations (see Example 13).


Thus, all in all, albeit W-symmetries can not be used to integrate an equation, at least in the usual way (see Section \ref{sec:straight} above), they are useful in mapping solutions to solutions.


\medskip\noindent
{\bf Example 11.} The equation
\beql{eq:exW1.1} dx \ = \ \a \, x \, dt \ + \ \b \, d w \eeq (where $\a, \b$ are real constants) admits, as seen above, the proper W-symmetry $X = x \pa_x + w \pa_w$. This acts by
\beq x \ \to \ \la \ x \ , \ \ \ w \ \to \la \ w \ \ \ \ (\la \in R) \ , \eeq hence maps our equation into the new equation
\beql{eq:exW1.2} d x \ = \ \a \, x \, dt \ + \ \la \, \b \, dw \ . \eeq
In this case, W-symmetry analysis just provides the trivial information that for a given realization $\wt{w} (t)$ of the driving process and a solution $\wt{x} (t)$ of the initial equation \eqref{eq:exW1.1} with initial datum $\wt{x}(0) = \wt{x}_0$, we also have a solution
$$ \^x (t) \ = \ \la \ \wt{x} (t) $$ with initial datum $\^x (0) = \^x_0 = \la \wt{x}_0$ for the transformed equation \eqref{eq:exW1.2}. This result is fully trivial: in fact, by writing
$$ x \ = \ \la \ \xi $$ the equation \eqref{eq:exW1.2} is mapped into \eqref{eq:exW1.1} (for the dynamical variable $\xi$). \EOE

\medskip\noindent
{\bf Example 12.}
The equation
\beql{eq:exW0.1} d x \ = \ a \, x \, dt \ + \ b \, \sqrt{x} \, d w \eeq
(where $a,b$ are nonzero real constants) admits the W-symmetry generated by $X = 2 x \pa_x + w \pa_w$; the action of this is a map (here $\la = \exp (s) \in R_+$)
$$ x \ \to \ \la \, x \ , \ \ \ w \to \la \, w \ . $$
It is a simple matter to check that this map just leaves invariant the equation \eqref{eq:exW0.1}: each of its three terms gets multiplied by the same factor $\la^2$.

In this case the W-symmetry tells us that if we have a solution $x(t)$ to this equation for a realization $w(t)$ of the driving process, we also have a solution $\^x (t) = \la^2 x(t)$ for a realization $\^w (t) = \la w(t)$ of the driving process.

We can of course also reinterpret the situation in the same terms as in Example 11 above; that is, consider -- together with \eqref{eq:exW0.1} -- the equation
\beql{eq:exW0.2} d \xi \ = \ a \, \xi \, dt \ + \ \la \, b \, \sqrt{\xi} \, d w \ . \eeq
Then for \emph{the same} realization of the driving process $w(t)$, any solution $x(t)$ to \eqref{eq:exW0.1} will correspond to a solution $\xi (t) = \la^2 x(t)$ to \eqref{eq:exW0.2}. \EOE

\medskip\noindent
{\bf Example 13.} Consider the $n$-dimensional system of proper Ito equations
\beql{eq:exW2.1} dx^i \ = \ - \, \a (\rho^2) \, x^i \, dt \ + \ \b (\rho^2 ) \, d w^i \ , \eeq
where we have written
$$ \rho^2 \ = \ \sum_{i=1}^n \( x^i \)^2 \ . $$
This admits $(i)$ a full $O(n)$ group of \emph{simultaneous} rotations in the $\xb$ and the $\wb$; and $(ii)$ \emph{only for $\a$ and $\b$ actually constant} the group of \emph{simultaneous homogeneous} dilations $\xb \to \la \xb$, $\wb \to \la \wb$.

In this case, applying the W-symmetries gives a trivial and a non fully trivial result:

\begin{itemize}
\item[$(i)$] The dilation symmetry permits -- in the case of linear equations, see above -- to relate solutions to the equation \eqref{eq:exW2.1} with a given initial condition $\xb_0$ and a given realization of the driving process $\wb (t)$ to solutions with initial condition $\^\xb_0 = \la \xb_0$ and the realization $\^\wb (t) = \la \wb (t)$ of the driving Wiener process; such a relation is analogous to those seen above and is quite trivial.
\item[$(ii)$] The rotation symmetry guarantees that once we have a solution $\xb (t)$ for a given initial condition $\xb_0$ and a given realization of the vector Wiener process $\wb (t)$, we also have a ``rotated'' solution for the ``rotated'' initial condition $\^\xb_0 = R \xb_0$ and the ``rotated'' realization of the Wiener process, $\^\wb (t) = R \wb (t)$; and this for any $R \in O(n)$. This fact is in a way also trivial, but can be recovered in symmetry terms \emph{only} by considering W-symmetries \gcite{GSW}.
\end{itemize}
{ $\ $ }  \EOE

\section{Discussion and conclusions}
\label{sec:discuss}

Our work is set within the framework of the symmetry approach for (Ito) stochastic differential equations \gcite{\sderef}, with a focus on integration of these by exploiting their symmetry properties \gcite{Koz1,Koz2,Koz3,Koz2018,GS17,GSW,Koz18a,Koz18b,GRQ1,GRQ2,GL1,GL2,GGPR}. We have noted that while the literature  dealing with \emph{deterministic standard symmetries} presents a  complete theory, the same cannot be said for \emph{random standard symmetries} \gcite{GS17,Koz18a,Koz18b} and for \emph{W-symmetries} \gcite{GSW}.

In particular we have noted that:
\begin{enumerate}
\item The Kozlov substitution related to a random standard symmetry will in general transform a proper Ito equation into a generalized Ito equation; thus if one aims at exploiting multiple symmetries -- when dealing with systems of Ito equations -- it is necessary to consider also symmetries of generalized Ito equations.
\item The determining equations for symmetry of these are (slightly) different from those for proper Ito equations; this requires to rework a number of results.
\item For random standard symmetries the standard proof showing that deterministic standard symmetries are preserved under change of variables does not apply directly (also for proper Ito equations), and one has to extend this proof to the random symmetry case, besides extending it to generalized Ito equations.
\item Related to the point above is the fact that the correspondence between symmetries of a given Ito equation and of the corresponding Stratonovich one has been firmly established (even for proper Ito and Stratonovich equations) for deterministic standard symmetries, and one has to extend that proof to the framework of random standard symmetries, besides extending it to generalized Ito and Stratonovich equation.
\item The latter task also requires to consider the Ito--Stratonovich map in the case of generalized Ito equations, which is not necessarily the same as for proper ones.
\item Besides standard symmetries, the literature also considers so called W-symmetries. These are essential in accounting for scaling and rotation symmetries involving both the dynamical variables and the ``noise variables'' modeled by Wiener processes \gcite{GSW}. But, while the Kozlov substitution allows to integrate equations possessing a standard symmetry, the use of W-symmetries with respect to integration is not studied in the literature.
\item An essential property for using W-symmetries in this direction would be their preservation under changes of variables (this is the equivalent for W-symmetries of the problems mentioned in items 3 and 4 above for random standard symmetries); but this has not been studied previously.
    \end{enumerate}

\noindent
In the present work we have considered the problems raised by the observations above. In particular, as a needed tool for our study we have derived the determining equations for symmetries of generalized Ito and Stratonovich equations. We have also determined the Ito--Stratonovich map in this case.

Our conclusions have been of different nature for the two classes of symmetries we have been considering:

\begin{itemize}
\item For random standard symmetries, we have shown that Kozlov theory is extended basically unharmed to this more general framework, both for proper and generalized Ito equations.
\item For W-symmetries there is nothing like this. In particular, they are not always preserved under changes of variables. Moreover they cannot be taken to the Kozlov form without affecting (too) deeply the nature of the equation under study; and the standard scaling form to which they can be brought -- at least for \emph{split} W-symmetries -- in a routine way are of no use towards integration.
\end{itemize}

\bigskip
\noindent
We hope our study, albeit limited to \emph{scalar} equations, will be of general use in the symmetry study of stochastic differential equations.

In particular, it should suggest to researchers dealing with (scalar or possibly $n$-dimensional systems of) Ito equation to disregard W-symmetries -- at least until a different use for them, e.g. along the lines mentioned in the footnote to Remark 9, is found.

On the other hand, our study should suggest to focus instead on standard symmetries, considering deterministic and random ones on the same level.

When considering systems of Ito equations and multiple reductions, we have shown that one should be ready to consider generalized Ito equations; but also that these can be dealt without hard problems -- at least from the symmetry viewpoint.

A number of topics, albeit related to our subject and quite natural in its light,  have remained outside our study. Thus e.g. we have not considered the special -- but physically most relevant -- case of equations issued by a (stochastic) variational problem \gcite{Yasue,Zam1,Zam2}, or the case of Ito equations arising from introducing stochastic terms in a ``second order'' equation \gcite{Koz2}.

Limitations to the scalar case has also allowed not to face problems which are inherent to the case of multiple symmetries and multiple symmetry reductions; in particular, how the Lie algebraic properties of the symmetry algebra would enter in multiple reductions. This problem is also present in the study of deterministic equations \gcite{\symmref}, and one should expect at least the same limitations as in the deterministic case.\footnote{See Example C.4 in Appendix \ref{app:higher} in this respect.}

\addcontentsline{toc}{subsection}{Acknowledgements}

\subsection*{Acknowledgements}

I am indebted to M.A. Rodr\'iguez for many discussions on symmetries of SDEs. The final version was improved thanks to questions raised by A.B. Cruzeiro at the SPT2023 workshop, and to remarks by an anonymous Referee. The work described in this paper was performed at SMRI, while on sabbatical leave from Universit\`a di Milano. My work is also supported by GNFM-INdAM and by INFN (sezione di Milano) through the MMNLP project.

\newpage

\begin{appendix}

\section{Derivation of the determining equations}
\label{app:deteqs}

In the main body of the text, we have given the determining equations without detail of their derivation. These will be given here, for both the Ito and the Stratonovich case. We will work directly with generalized Ito and Stratonovich equation;the proper case will be obtained by setting the $w$ derivatives of $f$, $b$, and $\s$ to zero; similarly, equations for standard symmetries will be obtained by setting $r=0$.

We will repeat some formulas already appearing in the main text, for ease of reference.

\subsection{Ito equations}

We consider a general vector field
\beql{app2:X} X \ = \ \vphi (x,t,w) \, \pa_x \ + \ r \, w \, \pa_w \ ; \eeq its infinitesimal action in$(x,t,w)$ space is given by
$$ x \ \to \ x \ + \ \eps \ \vphi (x,t,w) \ , \ \
t \ \to \ t  \ , \ \
w \ \to \ w \ + \ \eps \, r \, w \ . $$
We will consider the infinitesimal action of $X$ given by \eqref{app2:X} on a generalized Ito equation
\beql{app2:gIto} dx \ = \ f(x,t,w) \, dt \ + \ \s (x,t,w) \, dw \ . \eeq
At first order in $\eps$, and omitting the functional dependencies for ease of writing (it is understood that $f$, $\s$ and their derivatives are always evaluated in $(x,t,w)$), we have
\begin{eqnarray*}
dx \ + \ \eps \, d \vphi &=& \[ f \ + \ \eps \( \vphi \, f_x \ + \ r \, w \, f_w \) \] \, dt \\
& & \ + \ \[ \s \ + \ \eps \( \vphi \, \s_x \ + \ r \, w \, \s_w \) \] \ \( 1 \ + \ \eps \, r \) \, d w \\
&=& f \, dt \ + \ \s \, dw \\
& & \ + \ \eps \ \[ \( \vphi \, f_x \ + \ r \, w \, f_w \)  \, dt \ + \ \( \vphi \, \s_x \ + \ r \, w \, \s_w \ + \ r \, \s \) \, dw \] \ . \end{eqnarray*}
On solutions to \eqref{app2:gIto}, terms of order zero in $\eps$ cancel out, and we are left with
\beql{app2:dvphi1} d \vphi \ = \ \( \vphi \, f_x \ + \ r \, w \, f_w \)  \, dt \ + \ \( \vphi \, \s_x \ + \ r \, w \, \s_w \ + \ r \, \s \) \, dw \ . \eeq

We should now recall that in general
$$ d \vphi \ = \ \vphi_x \, dx \ + \ \vphi_t \, dt \ + \ \vphi_w \, dw \ + \ \frac12 \, \De (\vphi) \, dt \ ; $$ restricting again to solutions to \eqref{app2:gIto} we get
\beql{app2:dvphi2} d \vphi \ = \ \vphi_x \, \( f \, dt \ + \ \s \, dw \)  \ + \ \vphi_t \, dt \ + \ \vphi_w \, dw \ + \ \frac12 \, \De (\vphi) \, dt \ . \eeq

Putting together \eqref{app2:dvphi1} and \eqref{app2:dvphi2} we get
\begin{eqnarray*}
& & \( \vphi_t \ + \ f \, \vphi_x \ - \ \vphi \, f_x \ + \ \frac12 \, \De (\vphi) \ - \ r \, w \, f_w \) \ dt \\
& & \ + \ \( \vphi_w \ + \ \s \, \vphi_x \ - \ \vphi \, \s_x \ - \ r \, \s \ - \ r \, w \, \s_w \) \ dw \ = \ 0 \ . \end{eqnarray*}
As $dt$ and $dw$ are independent, hence their coefficients should vanish separately, we get the two equations
\begin{eqnarray}
\vphi_t \ + \ f \, \vphi_x \ - \ \vphi \, f_x \ + \ \frac12 \, \De (\vphi) & = & r \, w \, f_w \ , \\
\vphi_w \ + \ \s \, \vphi_x \ - \ \vphi \, \s_x \ - \ r \, \s & = & r \, w \, \s_w \ . \end{eqnarray}
These are exactly the determining equation \eqref{eq:gendeteq1}, \eqref{eq:gendeteq2} given in Section \ref{sec:ito}.

\medskip\noindent
{\bf Remark A.1.} In Section \ref{sec:ito}, we have also given eqs. \eqref{eq:fdenew1} and \eqref{eq:fdenew2}. These apply to transformations more general than those generated by $X$ in \eqref{app2:X}; in fact they refer to $X$ as in \eqref{eq:Xdefnew}. In order to obtain \eqref{eq:fdenew1}, \eqref{eq:fdenew2} we proceed as above; now (considering directly only terms of order $\eps$)
\beql{eq:newA1} d \vphi \ = \ (\vphi f_x + \xi f_w ) \, dt \ + \ (\vphi \s_x + \xi \s_w ) \, dw \ + \ \s \, d \xi \ . \eeq
Recalling that with the present general functional dependencies we have
\begin{eqnarray*} 
d \vphi &=& \vphi_t dt + \vphi_x dx + \vphi_w dw + \frac12 (\Delta \vphi ) dt \ , \\
d \xi &=& \xi_t dt + \xi_x dx + \xi_w dw + \frac12 (\Delta \xi ) dt \ , \end{eqnarray*}
and substituting in \eqref{eq:newA1} we obtain easily the equations \eqref{eq:fdenew1}, \eqref{eq:fdenew2}. \EOR

\subsection{Stratonovich equations}

We proceed in the same way for Stratonovich equations. Now the action of \eqref{app2:X} on the generalized Stratonovich equation
\beql{app2:gStrat} dx \ = \ b(x,t,w) \, dt \ + \ \s (x,t,w) \circ dw \eeq
is given, at first order in $\eps$ and with the same shorthand notation as above, by
\begin{eqnarray*}
dx \ + \ \eps \, d \vphi &=& \[ b \ + \ \eps \( b_x \, \vphi \ + \ b_w \, r \, w \) \] \, dt \\ & & \ + \ \[ \s \ + \ \eps \( \s_x \, \vphi \ + \ \s_w \, r \, w \ + \ r \, \s \) \] \, dw \ . \end{eqnarray*}
Terms of order zero in $\eps$ cancel out on solutions to \eqref{app2:gStrat}, and we are left with
\beql{app2:ps1} d \vphi \ = \ \( b_x \, \vphi \ + \ b_w \, r \, w \) \, dt \ + \ \( \s_x \, \vphi \ + \ \s_w \, r \, w \ + \ r \, \s \) \, dw \ . \eeq
On the other hand, recalling that now we should use standard calculus rather than Ito one, as we are working in the Stratonovich framework, we have
$$ d \vphi \ = \ \vphi_x \, dx \ + \ \vphi_t \, dt \ + \ \vphi_w \, dw \ ; $$ on solutions to \eqref{app2:gStrat} this reads
\beql{app2:ps2} d \vphi \ = \ \( \vphi_t \ + \ b \, \vphi_x \) \, dt \ + \ \( \vphi_w \ + \ \s \, \vphi_x \) \circ dw \ . \eeq
Putting together \eqref{app2:ps1} and \eqref{app2:ps2}, and recalling again that $dt$ and $dw$ are independent, we get the two equations
\begin{eqnarray}
\vphi_t \ + \ b \, \vphi_x \ - \ \vphi \, b_x  & = & r \, w \, b_w \ , \\
\vphi_w \ + \ \s \, \vphi_x \ - \ \vphi \, \s_x \ - \ r \, \s & = & r \, w \, \s_w \ . \end{eqnarray}
These are exactly the determining equation \eqref{eq:deteqstrat1}, \eqref{eq:deteqstrat2} given in Section \ref{sec:strat}.

\newpage

\section{The relation between generalized Ito versus Stratonovich equations}
\label{app:relationseqs}

In Section \ref{sec:relation} we have used formula \eqref{eq:stratform}, i.e. (repeating it here again for ease of reference)
\beql{app4:B} b \ = \ f \ - \ \frac12 \, (\s \, \s_x \ + \ \s_w ) \ , \eeq
providing the correspondence between a generalized Ito equation and the equivalent generalized Stratonovich equation.

We are now going to derive this formula through a simple computation; this just follows the steps of the usual derivation of the Ito-Stratonovich correspondence.

We consider three times $t_- < t_0 < t_+$, and correspondingly we will write $x_- = x(t_-)$, $x_0 = x(t_0)$, $x_+ = x(t_+)$; $w_- = w(t_-)$, $w_0 = w(t_0)$, $w_+ = w(t_+)$. We will also denote differences as follows:
\begin{eqnarray*}
dt = t_+ - t_- \ , & \de t_- = t_0 - t_- \ , & \de t_+ = t_+ - t_0 \ , \\
dx = x_+ - x_- \ , & \de x_- = x_0 - x_- \ , & \de x_+ = x_+ - x_0 \ , \\
dw = w_+ - w_- \ , & \de w_- = w_0 - w_- \ , & \de w_+ = w_+ - w_0 \ . \end{eqnarray*}

Let us now focus on the case where $dt = (t_+ - t_-)$ is an infinitesimal. In the (non-anticipating) Ito description we have
\beql{app4:ito} dx \ = \ f(x_-,t_-,w_-) \, dt \ + \ \s (x_-,t_-,w_-) \, dw \ . \eeq

On the other hand, from the Stratonovich point of view we obtain $x_\pm$ as
\beql{app4:stratx}
x_\pm \ = \ x_0 \ \pm \ \[ b (x_0,t_0,w_0) \, \de t_\pm \ + \ \s (x_0,t_0,w_0) \, \de w_\pm \] \ . \eeq
This entails\footnote{We are not using the notation $\s \circ dw$ because now we are dealing with real numbers, not with stochastic differential equations.}
\begin{eqnarray}
dx \ = \ x_+ \ - \ x_- &=& b(x_0,t_0,w_0) \, (\de t_+ + \de t_-) \ + \ \s (x_0,t_0,w_0) \, (\de w_+ + \de w_- ) \nonumber \\ &=& b(x_0,t_0,w_0) \, dt \ + \ \s (x_0,t_0,w_0) \, dw \ . \label{app4:stra} \end{eqnarray}

We should now compare \eqref{app4:ito} and \eqref{app4:stra}, i.e. the Ito and the Stratonovich determination of $dx$. However, in \eqref{app4:ito} the functions $f$ and $\s$ are expressed in terms of $(x_- ,t_- ,w_-)$, while in \eqref{app4:stra} the variables used are $(x,t,w)$. Using the assumption that $dt$ is an infinitesimal, we expand $f$ and $\s$ in a Taylor series:
\begin{eqnarray*}
f(x_-,t_-,w_-) &=& \[ f \ - \ \( \frac{\pa f}{\pa x} \, \de x_- \ + \ \frac{\pa f}{\pa t} \, \de t_- \ + \ \frac{\pa f }{\pa w} \, \de w_- \) \]_{(x_0,t_0,w_0)}\ , \\
\s(x_-,t_-,w_-) &=& \[ \s \ - \ \( \frac{\pa \s }{\pa x} \, \de x_- \ + \ \frac{\pa \s }{\pa t} \, \de t_- \ + \ \frac{\pa \s }{\pa w} \, \de w_- \) \]_{(x_0,t_0,w_0)} \ . \end{eqnarray*}
We can now proceed to compare \eqref{app4:ito} and \eqref{app4:stra}; all functions and derivatives are now evaluated in $(x_0,t_0,w_0)$. We get
\begin{eqnarray} b \, dt \ + \ \s \, dw &=& f \, dt \ + \ \s \, dw \ - \ \( \frac{\pa f}{\pa x} \, \de x_- \ + \ \frac{\pa f}{\pa t} \, \de t_- \ + \ \frac{\pa f}{\pa w} \, \de w_- \) \, dt \nonumber \\
& & \ - \ \( \frac{\pa \s}{\pa x} \, \de x_- \ + \ \frac{\pa \s}{\pa t} \, \de t_- \ + \ \frac{\pa \s}{\pa w} \, \de w_- \) \, dw \ . \end{eqnarray}
We can now cancel the term $\s dw$ appearing on both sides. Moreover, we should express $\de x_-$ in terms of $\de t_-$ and $\de w_-$. Using the Stratonovich description, we have
\beq \de x_- \ = \ b (x_0,t_0,w_0) \, \de t_- \ + \ \s (x_0,t_0,w_0) \, \de w_- \ . \eeq
Thus we obtain (again functions and derivatives are evaluated in $(x_0,t_0,w_0)$)
\begin{eqnarray*}
b \, dt &=& f \, dt \ - \ \( \frac{\pa f}{\pa x} \, (b \, \de t_- \ + \ \s \, \de w_- ) \ + \ \frac{\pa f}{\pa t} \, \de t_- \ + \ \frac{\pa f}{\pa w} \, \de w_- \) \, dt \\
& & \ - \ \( \frac{\pa \s}{\pa x} \, (b \, \de t_- \ + \ \s \, \de w_- ) \ + \ \frac{\pa \s}{\pa t} \, \de t_- \ + \ \frac{\pa \s}{\pa w} \, \de w_- \) \, dw \ . \end{eqnarray*}

We take now the limit $dt \to 0$, which also means $d w \to 0$. Discarding higher order infinitesimals, we get
\beq
b \, dt \ = \  f \, dt \ - \ \( \frac{\pa \s}{\pa x} \, \s \ + \ \frac{\pa \s}{\pa w} \) \, (\de w_- ) \, dw \eeq

Passing to consider expectations, we should recall $dw = \de w_+ + \de w_-$, and that $\de w_+$ and $\de w_-$ are independent. Thus we have
\beq \langle (\de w_-) \, \cdot \, (\de w_+ \ + \ \de w_- ) \rangle \ = \ \langle (\de w_- ) \cdot (\de w_+ ) \rangle \ + \ \langle (\de w_- )^2 \rangle \ = \ \langle (\de w_- )^2 \rangle \ = \ (\de t_- ) \ . \eeq
Note that so far $t_0$ can be \emph{any} point in the interval $(t_-,t_+)$ (such generalization of the Stratonovich approach is considered e.g. in \gcite{CST,GABT}). In the Stratonovich approach, everything is time-symmetric under time inversion, and $t_0 = (t_- + t_+)/2$ is the central point of the interval. With this choice, $\de t_- = \de t_+ = dt/2$, and we get
\beq
b \, dt \ = \  f \, dt \ - \ \frac12 \ \( \frac{\pa \s}{\pa x} \, \s \ + \ \frac{\pa \s}{\pa w} \) \, dt \ . \eeq
Again here all functions and derivatives are evaluated in $(x,t,w) = (x_0,t_0,w_0)$.

We have thus shown that for generalized Ito and Stratonovich equations, the noise terms are equal in both formulations, and the correspondence between their drift terms is exactly the one provided by \eqref{app4:B}.

\newpage

\section{The higher dimensional case}
\label{app:higher}

Our discussion has been confined to the one dimensional case, but the higher dimensional setting has been repeatedly mentioned as the motivation behind a number of the questions we discussed.

We will now have a glance at the higher dimensional setting; in particular, we briefly consider some simple (linear) examples to illustrate how the theory developed here is relevant in this higher dimensional context.

In this Appendix all indices will run from 1 to $n$ (the generalization to the case where the dimension of the $\xb = (x^1,...,x^n)$ and the $\wb = (w^1,...,w^n)$ vectors are not equal would be immediate), and summation over repeated indices will be understood.
We have to consider partial derivatives w.r.t. different $x^i$ and $w^k$ variables; our shorthand notation for these will be
\beql{appC.pd} \pa_i \ := \ \pa / \pa x^i \ , \ \ \ \^\pa_k \ := \ \pa / \pa w^k \ . \eeq

\subsection{The determining equations}
\label{app:deteqhigher}

We have, first of all, to write down the symmetry determining equations in the case of \emph{systems} of (proper or generalized) Ito equations
\beql{eq:Itosyst} d x^i \ = \ f^i (\xb , t , \wb ) \, dt \ + \ \s^i_{\ k} (\xb , t , \wb) \, d w^k \ . \eeq
In this case one proceeds exactly in the same way as shown in Appendix \ref{app:deteqs} and obtains, with the notation \eqref{appC.pd},
\begin{eqnarray}
\pa_t \vphi^i \ + \ f^j \, \pa_j \vphi^i \ - \ \vphi^j \, \pa_j f^i \ + \ \frac12 \, \De (\vphi^i ) &=& R^j_{\ \ell} \, w^\ell \, \^\pa_j f^i \ ,  \\
\^\pa_k \vphi^i \ + \ \s^j_{\ k} \, \pa_j \vphi^i \ - \ \vphi^j \, \pa_j \s^i_{\ k} \ - \ \s^i_{\ j} \, R^j_{\ k} &=& R^j_{\ \ell} \, w^\ell \, \^\pa_j \s^i_{\ k} \ . \end{eqnarray}
In this higher-dimensional case, the Ito Laplacian is given by
\beql{eq:DeltaN} \De (\phi) \ := \ \sum_{i,j=1}^n \[ \delta_{ij} \( \frac{\pa^2 \phi}{\pa w^i \pa w^j} \)  \ + \ 2 \, \s^i_{\ j} \ \( \frac{\pa^2 \phi}{\pa x^i \pa w^j} \) \ + \ \sum_{k=1}^n \s^i_{\ k} \s^j_{\ k} \(\frac{\pa^2 \phi}{\pa x^i \pa x^j} \) \] \ . \eeq

Similarly, for systems of Stratonovich equations
\beql{eq:Strasyst}  d x^i \ = \ b^i (\xb , t , \wb ) \, dt \ + \ \s^i_{\ k} (\xb , t , \wb) \circ d w^k \ , \eeq
one obtains
\begin{eqnarray}
\pa_t \vphi^i &+& b^j \, \pa_j \vphi^i \ - \ \vphi^j \, \pa_j b^i \ = \ R^j_{\ \ell} \, w^\ell \, \^\pa_j b^i \ , \label{AC.eq1} \\
\^\pa_k \vphi^i &+& \s^j_{\ k} \, \pa_j \vphi^i \ - \ \vphi^j \, \pa_j \s^i_{\ k} \ - \ \s^i_{\ j} \, R^j_{\ k} \ = \ R^j_{\ \ell} \, w^\ell \, \^\pa_j \s^i_{\ k} \ . \label{AC.eq2} \end{eqnarray}

Details of computations (in both cases) are left to the reader. Note that again the second set of determining equations is just the same in the Ito and in the Stratonovich cases.

\subsection{The Stratonovich map}

For the sake of completeness, we will also mention that -- again proceeding exactly as above, i.e. in this case as in Appendix \ref{app:relationseqs} -- the relation between a system of (possibly generalized) Ito equations \eqref{eq:Itosyst} and the corresponding system of (possibly generalized) Stratonovich equations \eqref{eq:Strasyst}
is obtained though the $n$-dimensional generalized Stratonovich map
\beql{app5:B} b^i \ = \ f^i \ - \ \frac12 \ \( \frac{\pa \s^i_k}{\pa x^j} \, \s^j_{\ m} \ + \ \frac{\pa \s^i_{\ k}}{\pa w^m} \) \, \de^{km} \ , \eeq
where summation over repeated indices is understood, and $\de^{km}$ is the Kronecker delta.

In order to prove this formula, we proceed nearly verbatim as in Appendix \ref{app:relationseqs}, except that we have to introduce indices, and consider vectors $\xb = (x^1,...,x^n)$ and $\wb = (w^1,...,w^n)$. Thus \eqref{app4:ito} and \eqref{app4:stratx} now read respectively
\begin{eqnarray}
dx^i &=& f^i (\xb_- , t_- , \wb_ ) \, dt \ + \ \s^i_{\ k} (\xb_- , t_- , \wb_- ) \, d w^k \ , \label{app5:ito} \\
x^i_\pm &=& x^i_0 \ \pm \ \[ b^i (\xb_0 , t_0 , \wb_0 ) \, \de t_\pm \ + \ \s^i_{\ k} (\xb_0 , t_0 , \wb_0 ) \, \de w^k_\pm \] \ . \label{app5:stratx} \end{eqnarray}
This entails
\begin{eqnarray}
dx^i \, = \, x^i_+ \, - \, x^i_- &=& b^i (\xb_0 , t_0 , \wb_0 ) \, \( \de t_+ + \de t_- \) \ + \ \s^i_{\ k} (\xb_0 , t_0 , \wb_0 ) \, \( \de w^k_+ + \de w^k_- \) \nonumber \\
&=& b^i (\xb_0 , t_0 , \wb_0 ) \, dt \ + \ \s^i_{\ k} (\xb_0 , t_0 , \wb_0 ) \, d w^k \ . \label{app5:strat} \end{eqnarray}

Again we have to compare \eqref{app5:ito} and \eqref{app5:strat} in the case where $dt$ is an infinitesimal, which we do by expanding $f^i$ and $\s^i_{\ k}$ as Taylor series. This yields
\begin{eqnarray*}
f^i(\xb_-,t_-,\wb_-) &=& \[ f^i \ - \ \( \frac{\pa f^i}{\pa x^j} \, \de x^j_- \ + \ \frac{\pa f^i}{\pa t} \, \de t_- \ + \ \frac{\pa f }{\pa w^m} \, \de w^m_- \) \]_{(\xb_0,t_0,\wb_0)}\ , \\
\s^i_{\ k} (\xb_-,t_-,\wb_-) &=& \[ \s^i_{\ k} \ - \ \( \frac{\pa \s^i_{\ k} }{\pa x^j} \, \de x^j_- \ + \ \frac{\pa \s^i_{\ k} }{\pa t} \, \de t_- \ + \ \frac{\pa \s^i_{\ k} }{\pa w^m} \, \de w^m_- \) \]_{(\xb_0,t_0,\wb_0)} \ . \end{eqnarray*}
Comparing \eqref{app4:ito} and \eqref{app4:stra} -- and simplifying notation as now all functions and derivatives are evaluated in $(\xb_0 , t_0 , \wb_0 )$ -- we get
\begin{eqnarray} b^i \, dt \ + \ \s^i_{\ k} \, dw^k &=& f^i \, dt \ + \ \s^i_{\ k} \, dw^k \nonumber \\
& & - \ \( \frac{\pa f^i}{\pa x^j} \, \de x^j_- \ + \ \frac{\pa f^i}{\pa t} \, \de t_- \ + \ \frac{\pa f^i}{\pa w^m} \, \de w^m_- \) \, dt \nonumber \\
& & - \ \( \frac{\pa \s^i_{\ k}}{\pa x^j} \, \de x^j_- \ + \ \frac{\pa \s^i_{\ k}}{\pa t} \, \de t_- \ + \ \frac{\pa \s^i_{\ k}}{\pa w^m} \, \de w^m_- \) \, dw^k \ . \end{eqnarray}
We can now cancel the term $\s^i_{\ k} dw^k$ appearing on both sides. Moreover, we should express $\de x^j_-$ in terms of $\de t_-$ and $\de w^m_-$. Using the Stratonovich description, we have
\beq \de x^j_- \ = \ b^j (\xb_0,t_0,\wb_0) \, \de t_- \ + \ \s^j_{\ m} (\xb_0,t_0,\wb_0) \, \de w^m_- \ . \eeq
Thus we obtain (again functions and derivatives are all evaluated in $(\xb_0,t_0,\wb_0)$)
\begin{eqnarray*}
b^i \, dt &=& f^i \, dt \ - \ \( \frac{\pa f^i}{\pa x^j} \, (b^j \, \de t_- \ + \ \s^j_{\ m} \, \de w^m_- ) \ + \ \frac{\pa f^i}{\pa t} \, \de t_- \ + \ \frac{\pa f^i}{\pa w^m} \, \de w^m_- \) \, dt \\
& & \ - \ \( \frac{\pa \s^i_{\ k}}{\pa x^j} \, (b^j \, \de t_- \ + \ \s^j_{\ m} \, \de w^m_- ) \ + \ \frac{\pa \s^i_{\ k}}{\pa t} \, \de t_- \ + \ \frac{\pa \s^i_{\ k}}{\pa w^m} \, \de w^m_- \) \, dw^k \ . \end{eqnarray*}

We take now the limit $dt \to 0$, which also means $d w^\ell \to 0$. Discarding higher order infinitesimals, we get
\beq
b^i \, dt \ = \  f^i \, dt \ - \ \( \frac{\pa \s^i_{\ k}}{\pa x^j} \, \s^j_{\ m} \ + \ \frac{\pa \s^i_{\ k}}{\pa w^m} \) \, (\de w^m_- ) \, dw^k \ . \eeq

Passing to consider expectations, we should recall $dw^\ell = \de w^\ell_+ + \de w^\ell_-$, and that $\de w^m_+$ and $\de w^m_-$ are independent. Thus we have
\begin{eqnarray}
\langle (\de w^m_- ) \, \cdot \, (d w^k ) \rangle &=& \langle (\de w^m_-) \, \cdot \, (\de w^k_+ \ + \ \de w^k_- ) \rangle \nonumber \\ & & \ = \ \langle (\de w^m_- ) \cdot (\de w^k_-) \ + \ \langle (\de w^m_- ) \cdot (\de w^k_+) \rangle \nonumber \\
& & \ = \ \langle (\de w^m_- ) \cdot (\de w^k_- ) \rangle \ = \ \de^m_{\ k} \ (\de t_- ) \ , \end{eqnarray}
where of course $\de^m_{\ k}$ is the Kronecker delta.

As in the discussion of the one-dimensional case given in Appendix \ref{app:relationseqs}, so far $t_0$ can be \emph{any} point in the interval $(t_-,t_+)$. In the Stratonovich approach we take $t_0 = (t_- + t_+)/2$; with this choice, $\de t_- = \de t_+ = dt/2$, and we get
\beq
b^i \, dt \ = \  f^i \, dt \ - \ \frac12 \ \( \frac{\pa \s^i_{\ k}}{\pa x^j} \, \s^j_{\ m} \ + \ \frac{\pa \s^i_{\ k}}{\pa w^m} \) \, \de^m_k \, dt \ . \eeq
Again here all functions and derivatives are evaluated in $(\xb,t,\wb) = (\xb_0,t_0,\wb_0)$.

We have thus shown that for generalized Ito and Stratonovich equations in arbitrary dimension $n$, the noise terms are equal in both formulations, and the correspondence between their drift terms is exactly the one provided by \eqref{app5:B}.

\subsection{Some two-dimensional examples}

We will now discuss some very simple two-dimensional examples, in order to illustrate how the theory discussed in the main text enters in the analysis of this framework. We will consider time-autonomous systems for the sake of simplicity; this will set a number of smooth functions of time to be actually real constants.

We adopt a simplified notation as we just have two dynamical variables and two driving processes; that is, we will denote the dynamical variables as $(x,y) = (x^1,x^2)$ and the driving Wiener processes as $(w,z) = (w^1,w^2)$.

\medskip\noindent
{\bf Example \ref{app:higher}.1} Consider the linear system
\begin{eqnarray}
dx &=& \( a_{11} \, x \ + \ a_{12} \, y \) \, dt \ + \ \( k_{111} \, x \ + \ k_{112} \, y \) \, d w \ + \ \( k_{121} \, x \ + \ k_{122} \, y \) dz \ , \nonumber \\ & & \label{exC1.dx} \\
dy &=& \( a_{21} \, x \ + \ a_{22} \, y \) \, dt \ + \ \( k_{211} \, x \ + \ k_{212} \, y \) \, d w \ + \ \( k_{221} \, x \ + \ k_{222} \, y \) dz \ , \nonumber \\ & & \label{exC1.dy} \end{eqnarray}
where $w$ and $z$ are independent Wiener processes, and $a_{ij}$, $k_{ij\ell}$ are numerical constants.

This system is obviously symmetric under the scaling vector field
\beq X \ = \ x \, \pa_x \ + \ y \, \pa_y \ ; \eeq a characteristic function for this vector field can be chosen as
$ z  :=  y/x $. We will thus pass to variables
\beql{exC1.etazeta} \eta \ := \ \log [|x|] \ , \ \ \zeta \ := \ y/x \ . \eeq
In the new variables, the symmetry vector field is just
$$ X \ = \ \pa_\eta \ . $$
The change of variables is singular in $x=0$, so let us restrict our attention to the half-plane $x > 0$.

We will choose
\beql{exC1:ksimp} k_{221} = 0 , \ k_{211} = 0 ; \ k_{212} = k_{111} , \  k_{121} = k_{222} \ , \eeq
and moreover
\beql{exC1:asimp} a_{21} = 0 ; \ a_{22} = a_{11} \ . \eeq
In this way the original system is written as
\begin{eqnarray}
dx &=& \( a_{11} \, x \ + \ a_{12} \, y \) \, dt \ + \ \( k_{111} \, x \ + \ k_{112} \, y \) \, dw \ + \ \( k_{222} \, x \ + \ k_{122} \, y \) \, dz  \ , \nonumber \\
dy &=& \( a_{11} \, y \) \, dt \ + \ \( k_{111} \, y \) \, dw \ + \ \( k_{222} \, y \) \, dz \ . \label{exC1.ssyst} \end{eqnarray}
It can easily be checked that this admits a second symmetry, given by
\beq Y \ = \ x \, \pa_y \ . \eeq
Actually, \eqref{exC1.ssyst} is the more general linear system admitting $Y$ as a second symmetry.

Note that the two symmetry vector fields commute,
$ [ X , Y ] = 0$; moreover, $X$ and $Y$ are both deterministic standard symmetries.

The expression of $Y$ in the new variables is
\beq Y \ = \ \zeta \, \pa_\eta \ - \ \zeta^2 \, \pa_\zeta \ = \ \zeta \ \( \pa_\eta \ - \ \zeta \, \pa_\zeta \) \ . \eeq
For $x$ and $y$ evolving according to \eqref{exC1.ssyst} the equations for the new dynamical variables are
\begin{eqnarray}
d \eta &=& \( a_{11} - \frac{k_{111}^2}{2} \, + \, \( a_{12} - k_{111} k_{112} \) \, \zeta \,   - \, \frac{k_{112}^2}{2} \,   \zeta^2 \) \, dt \nonumber \\
   & & \ + \ \( k_{111}+k_{112}
   \zeta \) \, d w \ + \ \( k_{222}+k_{122} \zeta \) \, d z \ , \\
d \zeta &=& \( \zeta
   \left(-k_{222}^2-k_{122} k_{222} \zeta
   + (k_{112}
   (k_{111}+k_{112} \zeta) \zeta  -a_{12})\right) \) \, dt \nonumber \\
   & & \ - \ \( k_{112}
   \zeta^2 \) \, d w \ - \ \( k_{122}
   \zeta^2 \) \, dz  \ . \end{eqnarray}

We should check if $Y$ is still a symmetry for these, i.e. if the determining equations \eqref{exC1.ssyst} are satisfied. This check amounts to a straightforward (and rather boring, hence omitted) computation, which shows this is indeed the case. Note that this had to be expected on the basis of our general results (actually, on the basis of previous results in the literature \gcite{GL1,GL2}), as we are dealing with \emph{deterministic standard symmetries}.

It may also be noted that in this system the r.h.s. does not depend on $\eta$ (which is indeed the purpose of changing variables); thus we have an autonomous equation for the dynamical variable $\zeta$. Note this has the symmetry $Y$, but depends on \emph{two} independent Wiener processes, so we are not guaranteed to be able to integrate it; see \gcite{GR22b} for a discussion of this framework. If we are able to get a solution $\zeta (t)$ for this equation, for a given realization of the Wiener processes $w(t)$ and $z(t)$, then the equation for $\eta (t)$ is also easily solved. \EOE

\medskip\noindent
{\bf Example \ref{app:higher}.2} A simpler -- but nontrivial -- situation is obtained by choosing, in the Example \ref{app:higher}.1 above,
$$ k_{112} \ = \ 0 \ . $$
We will also set, for ease of notation,
$$ a_{11} = \a , \ a_{12} = \b \ ; \ \ k_{111} = \kappa_1 , \ k_{122} = \kappa_2 , \ k_{222} = \kappa_3 \ . $$
In this way the original system reads
\begin{eqnarray*}
dx &=& \( \a \, x \ + \ \b \, y \) \, dt \ + \ \kappa_1 \, x \, dw \ + \ \( \kappa_3 \, x \ + \ \kappa_2 \, y \) \, dz \ , \\
dy &=& \a \, y \, dt \ + \ \kappa_1 \, y \, dw \ + \ \kappa_3 \, y \, dz \ ; \end{eqnarray*}
the equations for the new variables $(\eta , \zeta )$ are
\begin{eqnarray*}
d \zeta &=& \( \kappa_2 \kappa_3 - \b + \kappa_2^2 \zeta \) \, \zeta^2
 \, dt \ - \  \kappa_2 \, \zeta^2 \, dz \ , \\
d \eta &=& \( \a \, + \, \b \, \zeta \, - \, \frac{1}{2} ( \kappa_1^2 + \kappa_3^2 + 2 \kappa_2 \kappa_3 + \kappa_2^2 \zeta^2 ) \)  \, dt \nonumber \\
& & \ + \
 \kappa_1  \, d w \ + \ \( \kappa_3 + \kappa_2 \zeta \) \, d z  \ . \end{eqnarray*}
Now the equation for $\zeta (t)$ depends on a single Wiener process $z(t)$, and hence the symmetry under $Y = \zeta^2 \pa_\zeta$ allows to integrate this equation via the Kozlov substitution; the equation for $\eta$ is then a \emph{reconstruction equation}, which is readily integrated:
\begin{eqnarray} \eta (t) &=& \eta (0) \ + \ \int_0^t \( \a \, + \, \b \, \zeta (t) \, - \, \frac{1}{2} ( \kappa_1^2 + \kappa_3^2 + 2 \kappa_2 \kappa_3 + \kappa_2^2 \zeta^2 (t) ) \) \, dt \nonumber \\
   & & \ + \ \kappa_1 \, \int_0^t d w (\tau) \ + \ \int_0^t \( \kappa_3 + \kappa_2 \zeta (\tau) \) \, d z (\tau) \ . \label{exC2.eta} \end{eqnarray}

In fact, the Kozlov substitution yields the new variable
\beql{exC2.chi} \chi \ = \ - \, \int \frac{1}{\zeta^2} \ d \zeta \ = \ \zeta^{-1} \ , \eeq and Ito calculus yields in turn
$$ d \chi \ = \ ( \beta \ - \ \kappa_2 \, \kappa_3 ) \, dt \ + \ \kappa_2 \, dz \ . $$ This is promptly integrated, providing
$$ \chi (t) \ = \ \chi (0) \ + \ ( \beta \ - \ \kappa_2 \, \kappa_3 ) \, t \ + \ \kappa_2 \ \[ z (t) \ - \ z(0) \] \ ; $$
in order to obtain $\zeta (t)$ we should then just invert \eqref{exC2.chi} to get $\zeta (t) = 1 / \chi (t)$. At this point $\eta (t)$ is obtained through the formula \eqref{exC2.eta} above. \EOE

\medskip\noindent
{\bf Example \ref{app:higher}.3} (random standard symmetry) We will consider again linear systems of the general form \eqref{exC1.dx}, \eqref{exC1.dy}, thus admitting the scaling symmetry $X = x \pa_x + y \pa_y$. We will however choose coefficients so that the system also admits a second symmetry $Y$, which we want to be a random standard symmetry (the case where the second symmetry is a W-symmetry will be considered in the next Example).

We will consider the system
\begin{eqnarray}
dx &=& \frac12 \ \[ x \, dt \ + \ (x \, - \, y ) \, dw \ + \ (x \, + \, y ) \, dz \] \ , \label{exC3.dx} \\
dy &=& \frac12 \ \[ y \, dt \ + \ (y \, - \, x ) \, dw \ + \ (x \, + \, y ) \, dz \] \ , \label{exC3.dy} \ . \end{eqnarray}
This is homogeneous of degree one in $x$ and $y$, hence invariant under the scaling vector field
\beql{exC3.X} X \ = \ x \, \pa_x \ + \ y \, \pa_y \ ; \eeq
this system also admits a second symmetry vector field, given by
\beql{exC3.Y} Y \ = \ \( e^w \, + \, e^z \) \, \pa_x \ - \ \( e^w \, - \, e^z \) \, \pa_y \ . \eeq
Note that $X$ is a deterministic standard symmetry, while $Y$ is a random standard symmetry. The commutator gives
\beql{exC3.comm} \[ X , Y \] \ = \ - \, Y \ . \eeq

We change variables so to straighten the vector field $X$, i.e. pass to $(\eta , \zeta)$ defined in \eqref{exC1.etazeta} above. In the new variables,
\beq X \ = \ \pa_\eta \ , \ \ \ Y \ = \ e^{- \eta} \ \[ \( e^w \, + \, e^z \) \, \pa_\eta \ - \ \( (e^w - e^z ) \, + \, (e^w + e^z) \, \zeta \) \, \pa_\zeta \] \ . \eeq
The dynamical equations, which are standard Ito one, for the new variables are
\begin{eqnarray}
d \eta  &=& \frac14 \, (1 - \zeta^2) \, dt \ + \ \frac12 \, (1 - \zeta) \, dw \ + \ \frac12 \, (1 + \zeta) \, dz \ , \\
d \zeta &=& - \, \frac12 \, (1 - \zeta^2) \, \zeta \, dt \ - \ \frac12 \, (1 - \zeta^2) \, dw \ + \ \frac12 \, (1 - \zeta^2) \, dz \ . \end{eqnarray}
Note that $X$ is obviously still a symmetry for this system. Thus we have to solve the equation for $\zeta (t)$, which plays the role of the ``symmetry-reduced equation''; once we have solved this the equation for $\eta (t)$ is a -- promptly integrated -- reconstruction equation.
In this special case the noise coefficients for $dw$ and $dz$ are just the same up to a sign; thus we can perform a further change of variable and introduce
$$ \xi \ = \ \int \frac{- 2}{1 - \zeta^2} \ d \zeta \ = \ \log \[ \frac{\zeta - 1}{\zeta + 1} \] \ ; $$ by Ito calculus we get
$$ d \xi \ = \ dw \ - \ dz \ . $$

Our goal was, however, not to integrate equations \eqref{exC3.dx}, \eqref{exC3.dy}; but to investigate the fate of the second symmetry (in this setting, $Y$) when we reduce the equation using the first one (in this setting, $X$).

To this aim, we now have to check if the set of equations \eqref{AC.eq1}, \eqref{AC.eq2} hold with $x^1 = \eta$, $x^2 = \zeta$, $w^1 = w$, $w^2 = z$, and $f^i$, $\s^i_{\ j}$, $\vphi^i$ defined by the formulas above for $d \eta$, $d \zeta$, and $Y$. The check amounts to a straightforward computation. It turns out that all the equations \eqref{AC.eq2} are satisfied; as for \eqref{AC.eq1}, the equation for $i=1$ is satisfied, but the equation for $i=2$ -- which is our reduced equation, see above -- yields
$$ \frac14 \, e^{- \eta} \ \( e^w \, + \, e^z \) \ (1 \, - \, \zeta^2) \, \zeta \ = \ 0 \ , $$ hence is not satisfied.

We conclude that the symmetry $Y$ \emph{does not persist} under the change of variables $(x,y) \to (\eta , \zeta )$.

This appears to be in contradiction with our conclusions in Sect.\ref{sec:persW}; but the situation is actually a bit more complex. First of all, our discussion in Sect.\ref{sec:persW} only regarded \emph{scalar} equations; but this is not the problem, and indeed we know from other sources \gcite{GL1,GL2} that random standard symmetries should behave nicely under changes of variables also in higher dimension.

The problem here lies in the Lie-algebraic structure of the symmetry algebra. As we have noted above, the commutator is given by \eqref{exC3.comm}. Thus we expect, on the basis of general Lie-theoretic considerations (also holding for symmetries of deterministic equations \gcite{\symmref}) that while the $X$ symmetry is preserved when we operate a symmetry reduction under $Y$, the converse is not true. We have just seen this second part of the statement in action, so to conclude -- showing the first part of the statement in action for this example -- our discussion of the system \eqref{exC3.dx}, \eqref{exC3.dy} we should consider symmetry reduction under $Y$ and check that $X$ is still a symmetry for the reduced system.

We will thus not use \eqref{exC1.etazeta} but consider instead a \emph{different} change of variables, and introduce the new coordinates
\beql{exC3.xitheta}
\xi \ = \ \frac{x}{e^w + e^z} \ , \ \ \theta \ = \  \frac{(e^w - e^z) \, x \ + \ (e^w + e^z) \, y}{e^w + e^z} \ . \eeq
In these variables, the vector fields read
\beq X \ = \ \xi \, \pa_\xi \ + \ \theta \, \pa_\theta \ , \ \ \
Y \ = \ \pa_\xi \ . \eeq
The evolution equations for the new variables are
\begin{eqnarray}
d \xi &=& \frac12 \, \frac{e^w - e^z}{(e^w + e^z)^2} \, \theta \, dt \ - \ \frac12 \, \frac{\theta}{(e^w + e^z) } \, dw \ + \ \frac12 \, \frac{\theta}{(e^w + e^z) } \, dz \ , \\
d \theta &=& \frac12 \, \frac{(e^w - e^z)^2}{(e^w + e^z)^2} \, \theta \, dt \ + \ \frac{e^z }{(e^w + e^z) } \, \theta \, dw \ + \ \frac{e^w}{(e^w + e^z) } \, \theta \, dz \ . \end{eqnarray}
Note they are generalized Ito equations.

The r.h.s. of these do not depend on $\xi$ (hence $Y$ is a symmetry); the equation for $\theta (t)$ is the reduced equation, and if this is solved we promptly obtain $\xi (t)$, the equation for it being just a reconstruction equation. \EOE

\medskip\noindent
{\bf Example \ref{app:higher}.4} We now come to consider an Example with a W-symmetry. Let us take the system
\begin{eqnarray}
dx &=& \( \a \, x \ - \ \b y \) \, dt \ + \ a \, dw \ - \ b \, dz \ , \nonumber \\
dy &=& \( \b \, x \ + \ \a y \) \, dt \ + \ b \, dw \ + \ a \, dz \ . \label{exC4} \end{eqnarray}
It is immediate to check that this system admits the scaling symmetry
$$ X \ = \ x \, \pa_x \ + \ y \, \pa_y \ + \ w \, \pa_w \ + \ z \, \pa_z \ , $$
and the rotational symmetry
$$ Y \ = \ - y \, \pa_x \ + \ x \, \pa_y \ - \ z \, \pa_w \ + \ w \, \pa_z \ . $$
Both of these are W-symmetries, and they commute: $[X,Y] = 0$.

As discussed at length in the main text, see Sect.\ref{sec:Wscaling}, we cannot straighten these vector fields remaining within the class of (possibly generalized) Ito equations. We can, however, change variables so to put at least the dynamical variables part in a convenient form. In particular we can consider the change of variables suggested by the $X$ symmetry, i.e.
$$ (x,y) \ \to \ (\eta , \zeta ) \ , \ \ \eta \ = \ \log ( |x|) \ , \ \ \zeta \ = \ y/x \ . $$
In these coordinates -- that is, in the coordinates $(\eta,\zeta,w,z)$ -- we have
\begin{eqnarray}
X &=& \pa_\eta \ + \ w \, \pa_w \ + \ z \, \pa_z \ , \label{exC4.X} \\
Y &=& - \zeta \, \pa_\eta \ + \ (1 + \zeta^2) \, \pa_\zeta \ - \ z \, \pa_w \ + \ w \, \pa_z \ . \label{exC4.Y}
\end{eqnarray}

According to our discussion, see Sect.\ref{sec:persW}, we are not guaranteed in general that the W-symmetries $X$ and $Y$ are still symmetries after the change of variables. Note, in this regard, that here the noise coefficients do not depend on the dynamical variables $x$ and $y$, so that the discussion in Sect.\ref{sec:persW} -- in particular the Corollary to Lemma 1 -- would imply that $W$-symmetries are preserved under change of variables. But one should remember that the discussion given there (and in the papers cited there, e.g. \gcite{GSW}) does only apply to \emph{scalar} Ito equations, actually with a \emph{single} noise term. Thus one should explicitly check if symmetries survive the change of variables, and it is not clear what to expect \emph{apriori}.

The equations for the dynamical variables $\eta$ and $\zeta$ are
\begin{eqnarray}
d \eta  &=& \[ \(\a \, + \, \b \, \zeta \) \ - \ (1/2) \, e^{- 2 \eta} \, (a^2 + b^2) \] \, dt \nonumber \\
& & + \ e^{- \eta } \, \[ a \, d w \ + \ b \, dz \] \ , \label{exC4.deta} \\
d \zeta &=& \[ e^{- 2 \eta} \, (a^2 + b^2) \, \zeta \ - \ \b \, \( 1 \, + \, \zeta^2 \) \] \, dt \nonumber \\
& & \ - \ e^{- \eta} \, \[ (a \zeta + b) \, dw \ + \ (b \zeta - a) \, dz \] \ . \label{exC4.dzeta} \end{eqnarray}

Having the explicit form for the dynamical equations, i.e. \eqref{exC4.deta}, \eqref{exC4.dzeta}, and the explicit form for the vector fields, i.e. \eqref{exC4.X}, \eqref{exC4.Y}, we just have to verify if the determining equations are satisfied\footnote{It is a trivial remark -- but it may be worth recalling, to avoid any possible confusion -- that whenever we change variables, the Ito Laplacian $\De$ is defined using the noise coefficients for the equations of the new variables.}. Note we are not granted they are satisfied not only for the vector field $Y$, but also for the vector field $X$ which we have been (partially) straightening with the change of variables.

When we perform the required computations, it turns out that the determining equations are indeed satisfied, both for $X$ and for $Y$. \EOE

\medskip\noindent
{\bf Example \ref{app:higher}.5} We finally consider an example where the equations have non-constant noise coefficients and admit a W-symmetry. Note that (for scalar equations) while in the case of spatially constant noise coefficients we know the W-symmetries are preserved, our discussion of Sect.\ref{sec:persW} did not reach any definite conclusion in the case of noise coefficients depending on the dynamical (i.e. ``spatial'') variables. Thus we expect that, depending on concrete cases, W-symmetries can be preserved or lost when passing to adapted variables.

We consider the equations
\begin{eqnarray}
dx &=& (a_{11} \, x \ + \ a_{12} \, y ) \, dt \ + \ r \, k_1 \, x  \, dw \ + \ k_1 \, x \, dz \ , \nonumber \\
dy &=& (a_{21} \, x \ + \ a_{22} \, y ) \, dt \ + \ r \, (k_2 \, x \ + \ k_3  \, y) \, dw \ + \ (k_2 \, x \ + \ k_3 \, y) \, dz \ . \label{exC5}
\end{eqnarray}
Note this is degenerate, in the sense the matrices of the $r$ and the $\s$ coefficients are both degenerate.

Our system \eqref{exC5} admits as symmetries the scaling (now acting only on dynamical variables) vector field
\beq X \ = \ x \, \pa_x \ + \ y \, \pa_y \eeq
and the additional vector field
$$ Y_0 \ = \ w \, \pa_w \ - \ r \, z \, \pa_z \ . $$
It may be nicer to consider instead $Y = q X + Y_0$ with $Q$ an arbitrary constant; that is,
\beq Y \ = \ q \, x \, \pa_x \ + \ q \, y \, \pa_y \ + \ w \, \pa_w \ - \ r \, z \, \pa_z \ . \eeq
Straightening the field $X$, i.e. passing to the variables $(\eta , \zeta)$ as above, we get
\begin{eqnarray*}
d \eta &=& \( a_{11} \ - \ (k_1^2 / 2) \, (1 + r^2) \ + \ a_{12} \, \zeta  \) \, dt \ + \ k_1 \, r \, dw \ + \ k_1 \, dz \ , \\
d \zeta &=& \( a_{21} \ - \ k_1 \, k_2 \, (1 + r^2) \ + \ \( a_{22} - a_{11} + k_1 (k_1 - k_3) \) \, \zeta \ - \ a_{12} \, \zeta^2 \) \, dt \\
& & \ + \ r \, \( k_2 + (k_3 - k_1) \, \zeta \) \, dw \ + \ \( k_2 + (k_3 - k_1) \zeta \) \, d z \ . \end{eqnarray*}
The symmetry vector fields read, in these variables, as
$$ X \ = \ \pa_\eta \ , \ \ \ Y \ = \ q \, \pa_\eta \ + \ w \, \pa_w \ - \ r \, w \, \pa_z \ . $$

Again we just have to verify if the determining equations are satisfied. With straightforward computations, it turns out they are, both for $X$ and for $Y$. \EOE

\end{appendix}

\end{document}